\documentclass[prd,floatfix,superscriptaddress,showpacs,twocolumn,amssymb,amsmath,amsfonts,aps]{revtex4}
\usepackage{graphicx} 
\usepackage{dcolumn}  
\usepackage{longtable}
\usepackage{bm}       
\usepackage{subfigure}
\usepackage{multirow} %
\usepackage{sidecap} 
\usepackage{rotating}
\usepackage{longtable}
\usepackage{bbold}
\RequirePackage{xspace}

\def\Bbar    {\kern 0.18em\overline{\kern -0.18em B}{}\xspace}
\newcommand {\xcu} {X_{\{c,u\}}}
\def\ellm       {\ensuremath{\ell^-}\xspace}
\def\ellp       {\ensuremath{\ell^+}\xspace}
\newcommand {\barnuell} {\overline{\nu}_\ell}
\usepackage{relsize}
\def\babar{\mbox{\slshape B\kern-0.1em{\smaller A}\kern-0.1em
    B\kern-0.1em{\smaller A\kern-0.2em R}}}
\def\Vub  {\ensuremath{|V_{ub}|}\xspace}
\def\Vcb  {\ensuremath{|V_{cb}|}\xspace}
\newcommand {\epsR} {\epsilon_R}
\def\c     {\ensuremath{c}\xspace}
\def\q     {\ensuremath{q}\xspace}
\def\qbar  {\ensuremath{\overline q}\xspace}
\newcommand {\ctl} {\cos \theta_\ell}
\newcommand {\stl} {\sin \theta_\ell}

\newcommand {\ctv} {\cos \theta_V}
\newcommand {\stv} {\sin \theta_V}
\newcommand {\thetav} {\theta_V}
\def\qsq        {\ensuremath{q^2}\xspace}
\def\thetal     {\theta_\ell}
\def\thetaV     {\theta_V}
\def\GF   {G_F} 
\newcommand {\img} {\, Im}
\newcommand {\rel} {\, Re}
\def\BR         {{\ensuremath{\cal B}\xspace}}
\def\B       {\ensuremath{B}\xspace}

\def\hzsq        {\ensuremath{|H^L_0|^2}\xspace}
\def\hpsq        {\ensuremath{|H^L_+|^2}\xspace}
\def\hmsq        {\ensuremath{|H^L_-|^2}\xspace}
\def\ssq         {\ensuremath{|S^L|^2}\xspace}
\def\dzsq        {\ensuremath{|D^L_0|^2}\xspace}
\def\dpsq        {\ensuremath{|D^L_+|^2}\xspace}
\def\dmsq        {\ensuremath{|D^L_-|^2}\xspace}
\def\hpasq        {\ensuremath{|H^L_\parallel|^2}\xspace}
\def\hpesq        {\ensuremath{|H^L_\perp|^2}\xspace}
\def\dpasq        {\ensuremath{|D^L_\parallel|^2}\xspace}
\def\dpesq        {\ensuremath{|D^L_\perp|^2}\xspace}

\def\rhzdz       {\ensuremath{Re(H^L_0 D^{L\ast}_0)}\xspace}

\def\rshz        {\ensuremath{Re(S^L H_0^{L \ast})}\xspace}

\def\rsdz        {\ensuremath{Re(S^L D_0^{L \ast})}\xspace}

\def\hpdp       {\ensuremath{H^L_+ D^{L\ast}_+ + H^L_- D^{L\ast}_-}\xspace}

\def\rdpdmdz     {\ensuremath{Re((D^L_+ + D^L_-)D^{L\ast}_0)}\xspace}

\begin{document}
\author{Biplab Dey}
\affiliation{Physik-Institut, Universit\"{a}t Z\"{u}rich, CH-8057 Z\"{u}rich, Switzerland}
\date{\today}

\title{Angular analyses of exclusive $\Bbar\to X \ell_1 \ell_2$ with complex helicity amplitudes}

%
%
\begin{abstract}

We present the differential rates for exclusive $\Bbar\to X \ell_1 \ell_2$, where $\ell_1$ is a charged massless lepton and $\ell_2$ is a charged or neutral massless lepton, and $X$ is a mesonic system up to spin 2. The cases of interest are semileptonic (SL) $\Bbar\to \xcu\ellm \barnuell$ decays, and $\Bbar \to X_s \ellm \ellp$ where the the di-lepton can be $c \bar{c}$ resonances or non-resonant electroweak penguins (EWP). We consider helicity amplitudes having non-zero relative phases that can be potential new sources for CP-violation. Our motivations for these additional phases include a complex right-handed admixture in the hadronic weak charged current for the SL decays and complex Wilson coefficients in the effective Hamiltonians for the EWP decays. We demonstrate the efficacy of a novel technique of projecting out the individual angular moments in the full rate expression in a model-independent fashion. Our work is geared towards ongoing data analyses at $\babar$ and LHCb. 

\end{abstract} 

\maketitle

\section{Introduction}

The theory of semileptonic (SL) $B$ decays is a rich and well-studied subject~\cite{gilman_full_expr,richman_burchat_SL,korner,korner_schilcher,hagiwara_npb,hagiwara_plb}. Within the framework of the Standard Model (SM), this has been widely used to probe the nature of the electroweak interaction and the structure of the Cabibbo-Kobayashi-Maskawa (CKM) matrix. In particular, the CKM matrix elements $\Vcb$ and $\Vub$ can be extracted from the rates of the processes $\overline{B} \to \xcu \ellm \barnuell$, where $\xcu$ is an exclusive charm or charmless meson state, respectively.

\begin{figure}
\begin{center}
\subfigure[]{
{\includegraphics[width=1.5in]{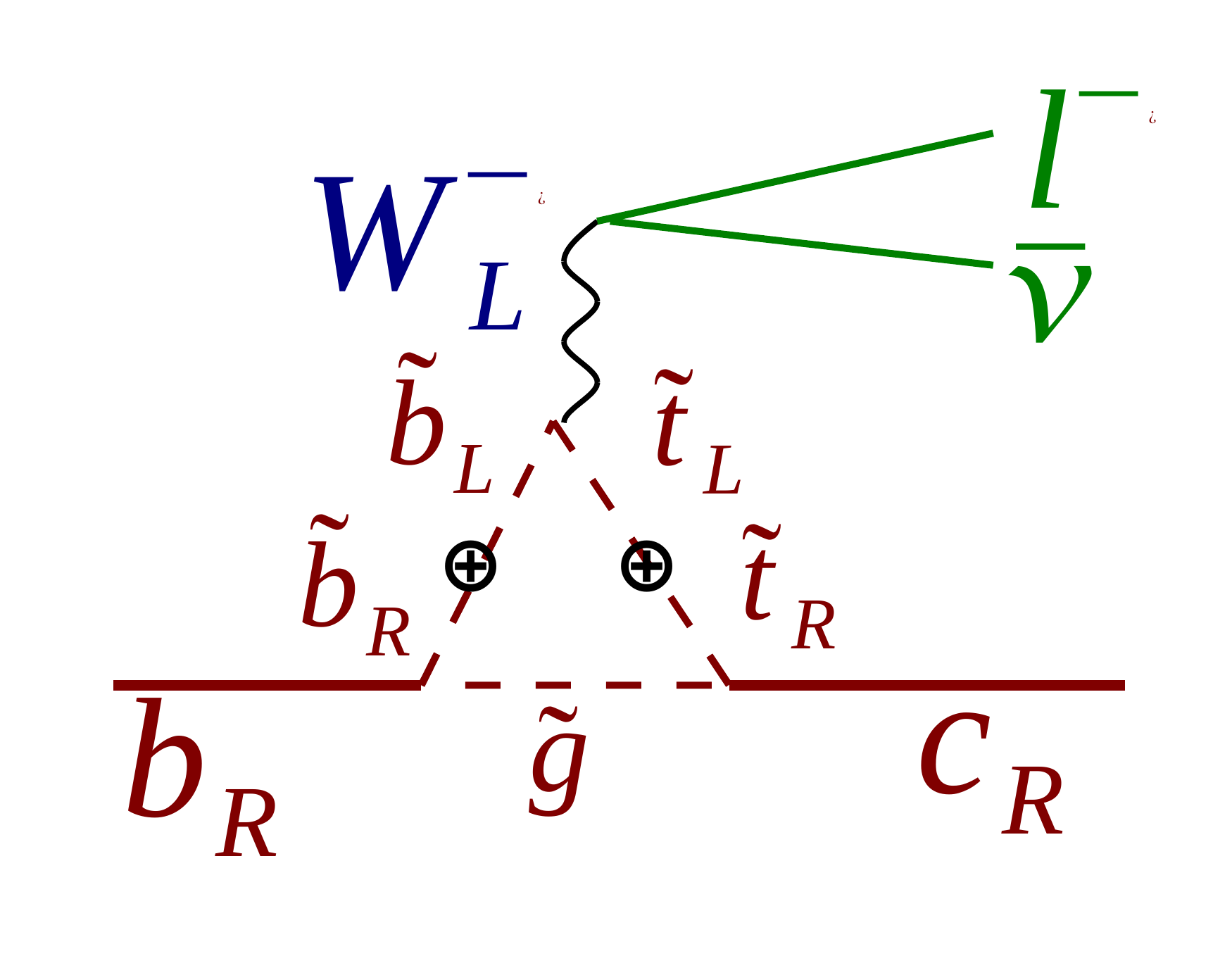}}}
\subfigure[]{
{\includegraphics[width=1.5in]{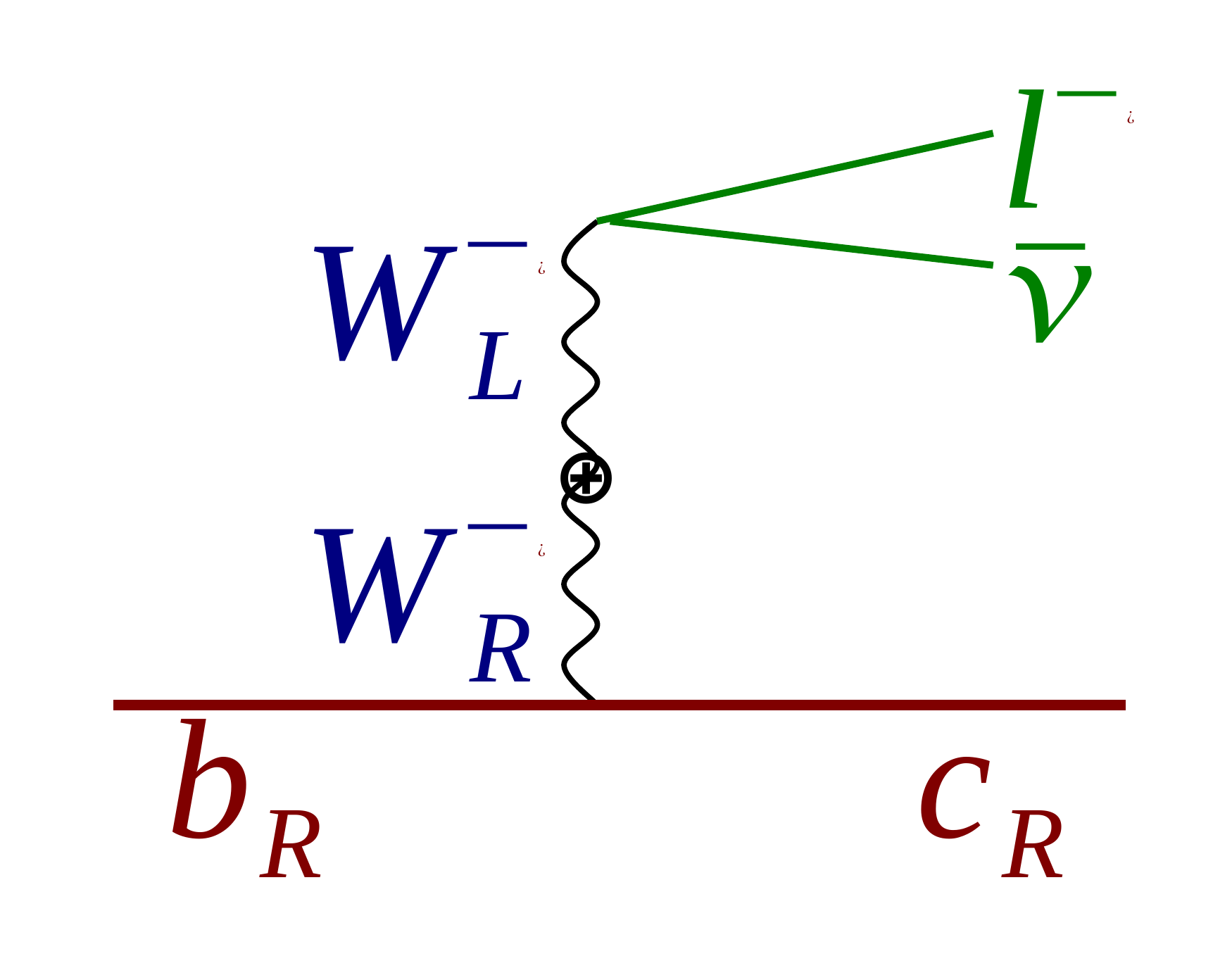}}}
\caption[]{\label{fig:rh_np} (Color online) Effective RH coupling arising in (a) gluino loop in supersymmetric models and (b) heavy RH $W$ boson in left-right symmetric models.}
\end{center}
\end{figure}

The full differential rate in the SM for these processes have been previously presented by several authors in Refs.~\cite{gilman_full_expr,richman_burchat_SL,korner,korner_schilcher,hagiwara_npb,hagiwara_plb}. The current article extends these results in the following fashion. Instead of assuming the relevant helicity amplitudes to be relatively real, as is the current status, we provide expressions corresponding to complex amplitudes. A specific motivation for admitting complex amplitudes in SL decays is to consider, instead of a purely left-handed (LH) weak charged current as in the SM, an additional complex right-handed (RH) admixture, $\epsR$, that could arise in new physics (NP) scenarios, as shown in Fig.~\ref{fig:rh_np}~\cite{crivellin2010,enomoto}. A complex non-zero $\epsR$ leads to additional angular terms in the full differential rate. In particular, a non-zero phase in $\epsR$ can lead to CP violation in SL $B$ decays~\cite{korner_schilcher,ng_prd_1997,ng_plb_1997_402, enomoto}. 

Consider on the other hand the process $\Bbar \to X \ellm \ellp$, where $X$ subsequently decays into two pseudoscalars, and the charged di-lepton system can be either be a $c \bar{c}$ resonance ($J/\psi$, $\psi(2S)$) or non-resonant electroweak penguins (EWP). It is well known that the helicity amplitudes here have non-zero relative phases~\cite{babar_verderi2005}. Compared to the SL case, where the leptonic current is purely LH, both LH and RH components exist for the charged di-lepton case. The LH and RH terms add incoherently to give the total rate. Therefore, while the number of angular observables remain the same, the number of independent real amplitude components to extract increases almost two-fold. The angular observables are not independent which leads to ambiguities in the solutions for the amplitudes~\cite{quim_complete_anatomy,ulrik_ambiguities}.

A simplification occurs for the case where the di-lepton is a resonant $\c\bar{c}$ meson that decays electromagnetically. Since electromagnetic interactions conserve parity, the LH and RH amplitudes are equal for this case. The reduced number of real amplitude components result in a single two-fold ambiguity, as explained in Sec.~\ref{sec:ambiguities}.
 
To sum up, in this article, we examine the generic $\Bbar \to X \ell_1 \ell_2$ decay, where $\ell_1$ is a charged massless lepton and $\ell_2$ is a charged or neutral massless lepton, and both the LH and RH helicity amplitudes can be non-zero, complex and independent of each other. The SL and the resonant $\c\bar{c}$ instances represent special cases leading to certain simplifications. We expand the full angular expression in an orthonormal basis of spherical harmonics and provide moments to project out each angular component. Since the basis is orthonormal, this reduces to a simple counting measurement. We explain how to extract the covariance matrix of the moments and the treatment of background subtraction, again, as counting measurements. As long as the set of basis functions is ``large-enough'', our method is the most model-independent way of describing the data, as inputs to theory modeling.

\section{The kinematic variables}

\begin{figure}
\centering
\includegraphics[width=2.5in]{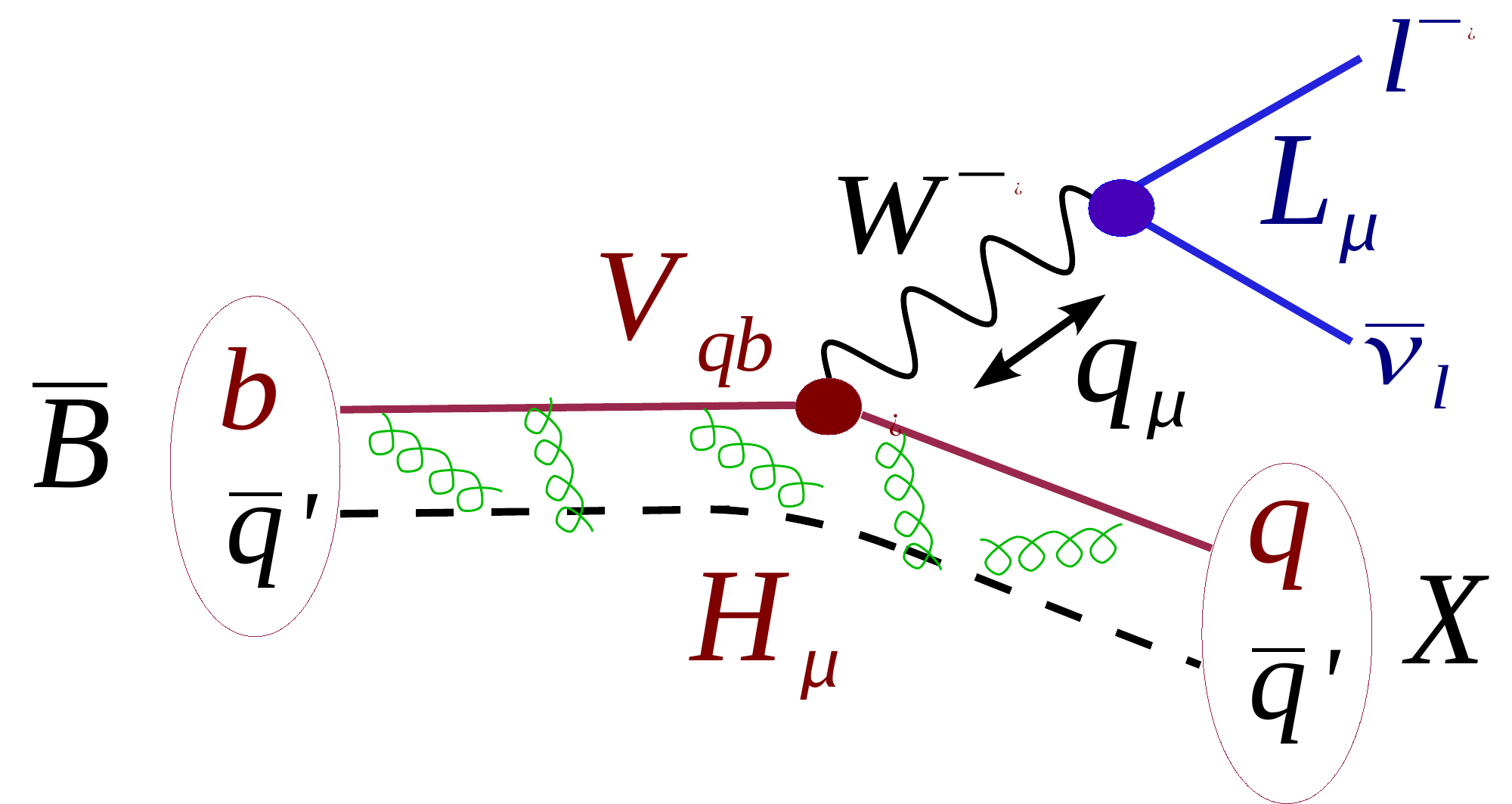}
\caption[]{\label{fig:SL_had_lep_factorization} (Color online) The quark level diagram for the SL decay $\Bbar(b\qbar')\to X(q \qbar ')\ellm \barnuell$ in the SM.}
\end{figure}

Consider the SL decay process $\Bbar(b\qbar')\to X(q\qbar ')\ellm \barnuell$ shown in Fig.~\ref{fig:SL_had_lep_factorization}. At the quark level, in the SM, this is a flavor changing process where a heavy $b$ quark emits a charged $W^\ast$ (off-shell) and decays into a lighter ${q\in \{u,c\}}$ quark, with the decay vertex containing the CKM matrix element $V_{qb}$. 
An important feature of SL $B$ decays is that the leptonic side interaction is well-understood, thereby facilitating study of the complicated non-perturbative QCD interactions that reside on the hadronic side. The momentum transfer squared between the leptonic and hadronic systems is $\qsq$. The hadronic side is thus probed by the $\qsq$ dependent form-factors (FF), just like in deep inelastic scattering (DIS), save that $\qsq >0$ is now timelike, instead of spacelike in the DIS case. For the EWP case, the $W^\ast$ can effectively be thought as being replaced by $\{\gamma^\ast, Z^\ast\}$. 

\subsection{Kinematics}
\label{sec:kinematics}

Without loss of generality, we take $\ell_1 \equiv \ell^-$ and $\ell_2 \equiv \barnuell$. We denote the 4-momenta of the parent $B$, the daughter meson $X$, the charged lepton $\ellm$ and $\barnuell$ as $p_B$, $p_X$, $p_\ell$ and $p_\nu$, respectively. The $W^\ast$ 4-momentum is $q = (p_B -p_X)$, so that
\begin{subequations}
\label{eqn:q2_w}
\begin{eqnarray}
q^2 &=& (p_B -p_X)^2 = m^2_B + m^2_X - 2m_B E_X\\
w &\equiv& v_B \cdot v_X = \frac{p_B}{m_B} \cdot \frac{p_X}{m_X} = \frac{E_X}{m_X} \nonumber \\ &=& \frac{m^2_B + m^2_X - q^2}{2m_B m_X},
\end{eqnarray}
\end{subequations}
where $E_X$ is the energy and $w$ is the $\gamma$-factor of the $X$ as seen in the $B$ rest frame (RF). If we consider the breakup of $B\to X W^\ast$ as a two-body decay, where the virtual $W^\ast$ boson has mass $\sqrt{\qsq}$, the two-body breakup momentum is given by
\begin{equation}
\label{eqn:k_expr}
{\bf k} = \sqrt{ \frac{(m^2_B-q^2 +m_X^2)^2}{4m^2_B} -m^2_X }.
\end{equation}
Two kinematic limits are of special interest. At ``zero-recoil'', ${\bf k} = 0$ and the $W^\ast$ attains the maximum allowed virtual mass, $\sqrt{\qsq} = \sqrt{\qsq_{\mbox{\scriptsize max}}} = (m_B -m_X)$. Since the meson $X$ is at rest in the $B$ RF now, the $\gamma$-factor $w = w_{\mbox{\scriptsize min}} = 1$. This kinematic region is convenient for lattice and heavy quark effective theory calculations. On the other hand, at $\qsq = \qsq_{\mbox{\scriptsize min}} \approx 0$ (for the massless leptons), the breakup momentum is largest 
\begin{equation}
{\bf k} _{\mbox{\scriptsize max}} = \frac{m^2_B -m^2_X}{2m_B}.
\end{equation}
Since the breakup momentum and the $\gamma$-factor $w$ are related as
\begin{equation}
{\bf k} = m_X\sqrt{w^2-1}.
\end{equation}
we also have
\begin{equation}
w_{\mbox{\scriptsize max}} = \frac{m^2_B +m^2_X}{2m_B m_X},
\end{equation}
or, the $\beta$-factor as
\begin{equation}
\beta_{\mbox{\scriptsize max}} = \frac{m^2_B - m^2_X}{m^2_B + m^2_X}.
\end{equation}
This ``large-recoil'' region is convenient for light-cone sum rules and soft collinear effective theory calculations.

\begin{figure}
\centering
\includegraphics[width=3in]{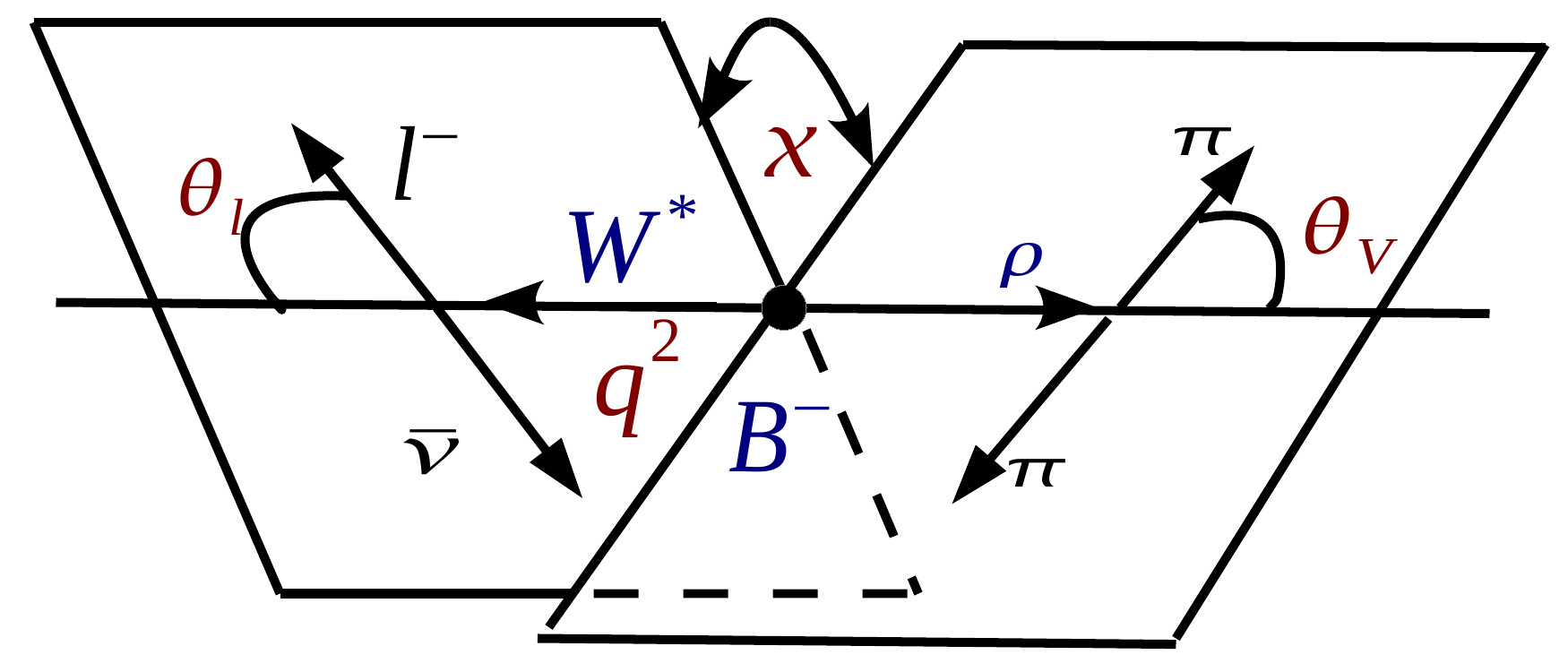}
\caption[]{\label{fig:btorhoellnu_angle_vars} (Color online) The set of four kinematic variables ${\phi \in \{q^2,\chi,\ctl, \ctv\}}$ for the SL decay chain ${\Bbar \to \rho (\to \pi \pi) W^\ast (\to \ellm \barnuell)}$.}
\end{figure}

\begin{figure}[h]
\begin{center}
\subfigure[]{
{\includegraphics[width=2.85in]{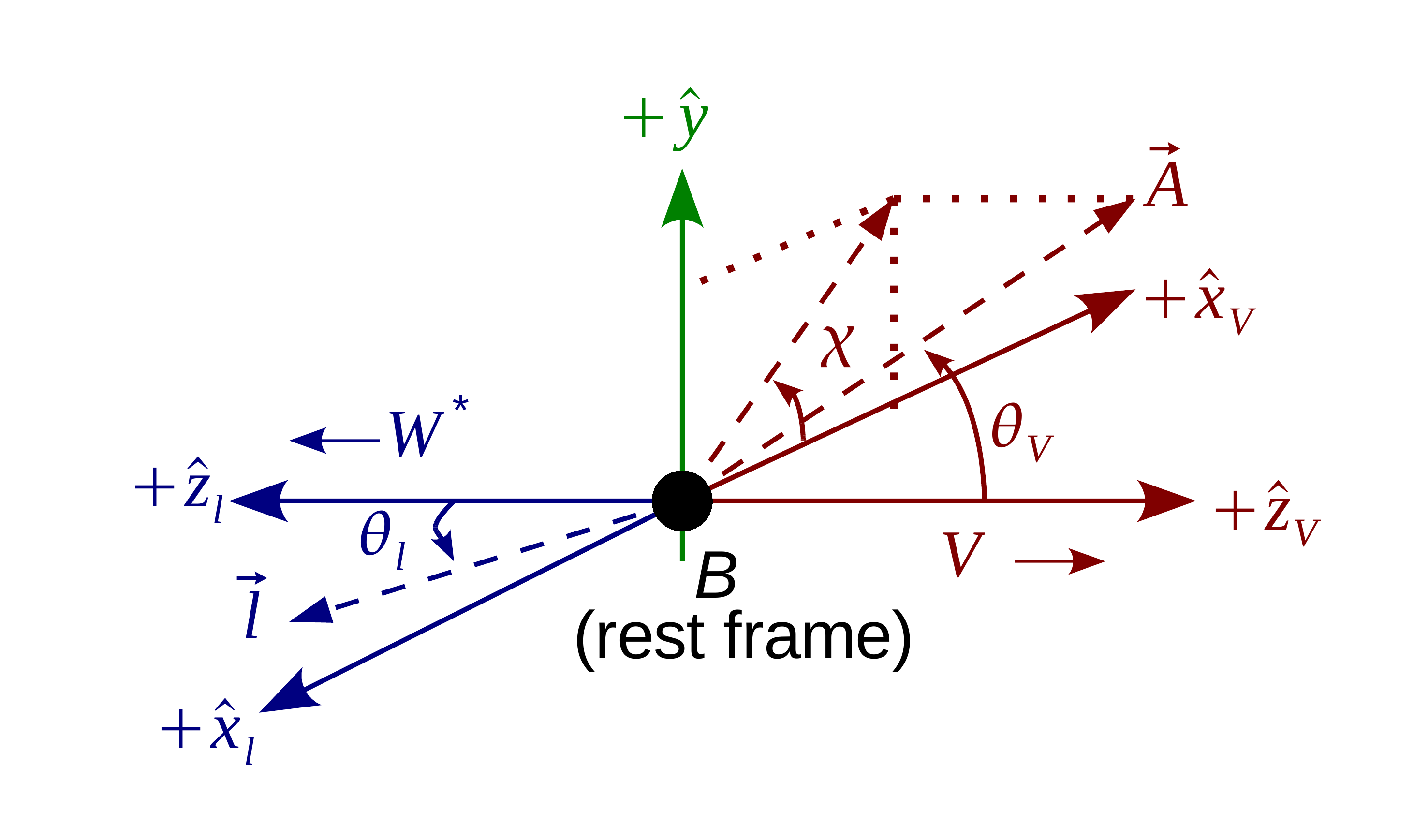}}
}
\subfigure[]{
{\includegraphics[width=2.85in]{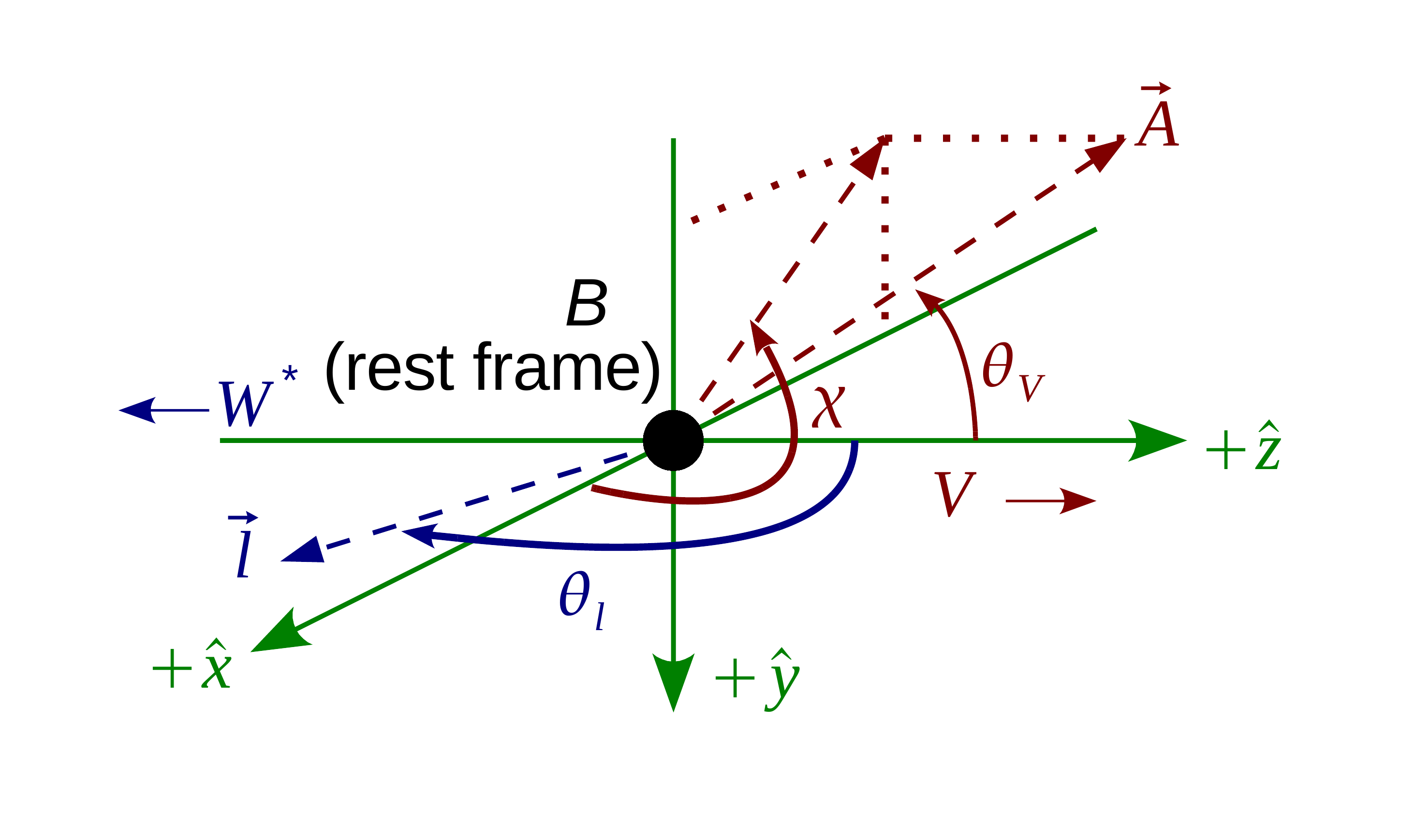}}
}
\end{center}
\caption{\label{fig:helicity_angles} (Color online) The reference frames for calculating $\thetal$, $\thetav$ and $\chi$ for $\Bbar \to V W^\ast (\to \ellm \barnuell)$ conforming to (a) Fig.~2 in Gilman-Singleton~\cite{gilman_full_expr} and (b) Fig.~3 in Hagiwara~\cite{hagiwara_npb}. The relations between the two sets of angles are given in the text.}
\end{figure}

When the outgoing meson in a vector meson, its polarization is important as well. The vector meson decay products act as the analyzer. For example, in the case of ${\rho \to \pi \pi}$ shown in Fig.~\ref{fig:btorhoellnu_angle_vars}, the analyzer ($\hat{A}$) is the $\pi$ momentum direction in the $\rho$ helicity frame with respect to the $\Bbar$ RF. This defines the helicity angle $\thetaV$. For ${\omega \to \pi^+ \pi^- \pi^0}$, the normal to the $\omega$ decay plane plays the role of the analyzer. The last additional kinematic variable is $\chi$, the dihedral angle between the $W^\ast$ and the vector meson decay planes in the mother $\Bbar$ RF and care must be taken to note the quadrant of the angle $\chi$ (see Fig.~\ref{fig:helicity_angles}). We refer to the set of four kinematic variables as $\phi \equiv \{\qsq, \ctl, \ctv, \chi\}$. 




\subsection{Sign conventions of $\{\thetal,\thetaV,\chi\}$ for $\Bbar \to X \ell_1 \ell_2$}
\label{sec:sign_conventions}

We stress here that the angles in this section are for the $\Bbar$ decay (that is, a $b$ quark transition). The CP conjugate case is decsribed in Sec.~\ref{sec:cp_conj}.

\subsubsection{SL case: $\{\ell_1, \ell_2 \}\equiv \{ \ell^-, \barnuell\}$}

We follow the definition of the angles in Fig.~2 of Gilman-Singleton~\cite{gilman_full_expr}. We first boost everything to the $\Bbar$ RF. There are two sets of co-ordinate axes, $\{\hat{x}_\ell,\hat{y}_\ell,\hat{z}_\ell\}$ and $\{\hat{x}_V,\hat{y}_V,\hat{z}_V\}$, as shown in Fig.~\ref{fig:helicity_angles}a for the vector meson (V) case. These are the helicity frames of the $W^\ast$ and the $V$. The connection is that $\hat{x}_\ell \equiv -\hat{x}_V$, $\hat{y}_\ell \equiv \hat{y}_V$ and $\hat{z}_\ell \equiv -\hat{z}_V$. The dihedral angle $\chi = \varphi_\ell + \varphi_V$, where we note that the azimuthal angles $\varphi_\ell$ and $\varphi_V$ are calculated in two different frames. We set $\varphi_\ell = 0$ by ensuring that the charged lepton $\ell$ lies in the $\hat{x}_\ell$-$\hat{z}_\ell$ plane and and has the $x$-component of its momentum $>0$. This completely fixes the quadrant of $\chi$, and therefore the signs of its sine and cosine. 

To measure $\theta_\ell$ and $\theta_V$, we boost to the $W^\ast$ and $V$ rest frames and measure the polar angles of the $\ell$ and $\vec{A}$, respectively. Here $\vec{A}$ is the analyzing direction of the vector meson decay as tabulated in Table~\ref{table:anal_dir}.

Korner-Schuler~\cite{korner} and Hagiwara~\cite{hagiwara_plb, hagiwara_npb} follow a different convention where both the orientations of the axes for the leptonic and hadronic systems are the same. The relations are
\begin{subequations} 
\label{eqn:gs2ks}
\begin{eqnarray}
\theta^{\mbox{\scriptsize KS}}_\ell &\equiv& \pi - \theta^{\mbox{\scriptsize GS}}_\ell \\
\theta^{\mbox{\scriptsize KS}}_V &\equiv& \theta^{\mbox{\scriptsize GS}}_V \\
\chi^{\mbox{\scriptsize KS}} &\equiv& \pi + \chi^{\mbox{\scriptsize GS}}
\end{eqnarray}
\end{subequations}
where the ``KS'' superscript refers to Korner-Schuler/Hagiwara and the ``GS'' superscript to Gilman-Singleton. 

The conventions followed by Richman-Burchat~\cite{richman_burchat_SL} (``RB'') on the other hand are related to the GS definitions as
\begin{subequations} 
\label{eqn:gs2rb}
\begin{eqnarray}
\theta^{\mbox{\scriptsize RB}}_\ell &\equiv& \theta^{\mbox{\scriptsize GS}}_\ell \\
\theta^{\mbox{\scriptsize RB}}_V &\equiv& \theta^{\mbox{\scriptsize GS}}_V \\
\chi^{\mbox{\scriptsize RB}} &\equiv& \pi + \chi^{\mbox{\scriptsize GS}}
\end{eqnarray}
\end{subequations}

We adhere to the GS conventions in this work.


\subsubsection{EWP case: $\{\ell_1, \ell_2 \}\equiv \{ \ell^-, \ell^+ \}$}

We again follow the GS conventions with the single replacement $\barnuell \to \ell^+$ everywhere. Compared to other EWP conventions in the literature~\cite{ulrik_ambiguities,kruger_matias,luwang,altman_jhep09,bobeth_jhep08,melikhov98}, the only change is
\begin{align}
\label{eqn:thetal_flip}
\theta^{\mbox{\scriptsize GS}}_\ell &\equiv \pi - \theta^{\mbox{\scriptsize EWP}}_\ell,
\end{align}
where the superscript ``EWP'' refers to the aforementioned theory references (see also appendix).

\begin{table}
\begin{center}
\begin{tabular}{ c|c  } 
Decay topology& $\vec{A}$ direction \\ \hline \hline 
$D^\ast \to D \pi$ & $\vec{p}_D$ \\ \hline
$D^\ast \to D \gamma$ & $\vec{p}_D$  \\ \hline
$\rho^\pm/f_0 \to \pi^\pm \pi^0$ & $\vec{p}_{\pi^0}$  \\ \hline
$\rho^0/f_0 \to \pi^+ \pi^-$ & $\vec{p}_{\pi^+}$ \\ \hline
$\omega \to \pi^+ \pi^- \pi^0$ & $\vec{p}_{\pi^+} \times \vec{p}_{\pi^-}$\\ \hline
$K^\ast \to K\pi$ & $\vec{p}_{K}$
\end{tabular}
\caption{The analyzing direction $\vec{A}$ in Fig.~\ref{fig:helicity_angles} for the different mesons in a $\Bbar$ decay. See also Sec.~\ref{sec:cp_conj} for the $B$ case.}
\label{table:anal_dir}
\end{center}
\end{table}

\section{Effective Hamiltonians}

\subsection{SL decays}

Consider the process $b \to q \ellm \barnuell$ (where $q \in \{c, u\}$ and $\ell \in \{e, \mu \}$) in terms of an effective 4-Fermi interaction Hamiltonian:
\begin{align}
\label{eqn:eff_hamiltonian} 
\mathcal{H_{\mbox{\scriptsize eff} }} &=  \frac{2 \GF V^L_{qb}}{\sqrt{2}} \left[ \left( g^V \bar{q} \gamma_\mu b - g^A \bar{q} \gamma_\mu \gamma_5 b\right) \bar{\ell} \gamma^\mu \nu_L \right. \nonumber \\
&\qquad \left. \hspace{0.1cm}+ \bar{q} \left (g^S + g^P \gamma_5\right)b \bar{\ell} \nu_L \right],
\end{align}
where we have assumed only LH neutrinos ($\nu_L = \frac{1-\gamma_5}{2} \nu$) and neglected any tensor terms associated with baryon and lepton number violations (leptoquark models~\cite{jongphillee_tensor}). Here, $V^L_{qb} \equiv V_{qb} $ denotes the usual LH CKM matrix element in the SM. The vector and the axial interactions are written as $g_V = (1 + \epsR)$ and $g_A = (1 - \epsR)$ and in general, $\epsR$ is allowed to be complex to incorporate additional CP violating effects~\cite{korner_schilcher,ng_prd_1997,ng_plb_1997_402, enomoto}. There are also two terms, $g^S$ and $g^P$, corresponding to scalar and pseudoscalar interactions, respectively. To retrieve the SM part, one puts ${\epsR = g^S = g^P = 0}$.

\subsubsection{The $\bar{B} \to S \ellm \bar{\nu}_\ell$ case}

The transition matrix element pertaining to the process $\Bbar \to S \ellm \barnuell$, where $S$ is a a $0^+$ scalar state, is then $\mathcal{M}_S = \langle S|\mathcal{H_{\mbox{\scriptsize eff} }} |\Bbar\rangle$. We note here that a negatively charged lepton and a RH anti-neutrino is being produced (since we have allowed for extra phases, we have to be careful about CP conjugates now). The hadronic matrix elements corresponding to the terms $g^{\{V,A,P,S\}}$ are written in terms of two form factors $u_+(\qsq)$ and $u_0(\qsq)$~\cite{chen_jhep2006,kln_b2pipilnu}:
\begin{subequations}
\label{eqn:full_ffs_S}
\begin{align}
\langle S |\bar{q} \gamma_\mu b|\Bbar\rangle_V  &= 0 \\
\langle S |\bar{q} \gamma_\mu \gamma_5b|\Bbar\rangle_A  &= u_+(\qsq) \Big((p_B + p_S)_\mu  - \frac{ (p_B + p_S)\cdot q }{\qsq} q_\mu \Big) \nonumber \\ 
& \hspace{1.2cm}+ u_0(\qsq) \frac{ (p_B + p_S)\cdot q }{\qsq} q_\mu \\
\langle S |\bar{q}b|\Bbar\rangle_S  &= 0 \\
\langle S |\bar{q} \gamma_5b|\Bbar\rangle_P  &\approx u_0(\qsq) \frac{m^2_B - m^2_S}{m_b + m_q}.
\end{align}
\end{subequations}
Since parity factors multiply, the right hand side in Eq.~\ref{eqn:full_ffs_S}a has to be an axial vector, which one can not construct out of the two vectors $p_S$ and $p_B$. Therefore only the $g^A$ term survives in Eq.~\ref{eqn:full_ffs_S}b, while the $g^V$ term in Eq.~\ref{eqn:full_ffs_S}a is zero. Eq.~\ref{eqn:full_ffs_S}b has been written in a form that is non-singular at $\qsq \to 0$. However, for the light leptons, the $q^\mu$ terms in go to zero when dotted with the leptonic charged current $\bar{\ell} \gamma^\mu \nu_L$. This can be seen by using $q = (p_\ell + p_\nu)$ and the Dirac equation for the (massless) leptons. Hence, all terms proprtional to $q_\mu$ can be dropped. Eq.~\ref{eqn:full_ffs_S}c and Eq.~\ref{eqn:full_ffs_S}d follow from Eq.~\ref{eqn:full_ffs_S}b and Eq.~\ref{eqn:full_ffs_S}b, respectively, by dotting with $q_\mu$ and invoking the Dirac equation at the quark level. In all, the transition matrix element reads: 
\begin{align}
\label{eqn:full_amp_S}
\mathcal{M}_S &= \frac{2 \GF V_{qb}}{\sqrt{2}} \Big\{ g_A  u_+(\qsq) (p_B + p_S)_\mu \bar{\ell}\gamma^\mu \nu_L \nonumber \\
& \hspace{2cm}+ g_P u_0(\qsq) \frac{m^2_B - m^2_S}{m_b + m_q}  \bar{\ell} \nu_L \Big\}
\end{align}
As we will see later, the $g_P$ term can be ignored for the massless lepton case, and only the $g_A$ term will remain.

\subsubsection{The $\bar{B} \to P \ellm \bar{\nu}_\ell$ case}

When the outgoing meson is a $0^-$ pseudoscalar state $P$, following the argument given above, the $g_A$ term vanishes and there is only a $g_V$ contribution, with the two form factors $f_+(\qsq)$ and $f_0(\qsq)$: 
\begin{subequations}
\label{eqn:full_ffs_P}
\begin{align}
\langle P |\bar{q} \gamma_\mu b|\Bbar\rangle_V  &= f_+(\qsq) \left((p_B + p_P)_\mu  - \frac{ (p_B + p_P)\cdot q }{\qsq} q_\mu \right) \nonumber \\ 
& \hspace{2cm} + f_0(\qsq) \frac{ (p_B + p_P)\cdot q }{\qsq} q_\mu \\
\langle P |\bar{q} \gamma_\mu \gamma_5 b|\Bbar\rangle_A  &= 0 \\
\langle P |\bar{q} \gamma_5 b|\Bbar\rangle_P  &= 0 \\
\langle P |\bar{q} b|\Bbar\rangle_S  &\approx f_0(\qsq) \frac{m^2_B - m^2_P}{m_b + m_q},
\end{align}
\end{subequations}
and the amplitude reads:
\begin{align}
\label{eqn:full_amp_P}
\mathcal{M}_P &= \frac{2 \GF V_{qb}}{\sqrt{2}} \Big\{ g_V  f_+(\qsq) (p_B + p_P)_\mu \bar{\ell}\gamma^\mu \nu_L \nonumber \\
& \hspace{2cm}+ g_S f_0(\qsq) \frac{m^2_B - m^2_P}{m_b + m_q}  \bar{\ell} \nu_L \Big\}.
\end{align}
As in the scalar case, the $g_S$ can be ignored for the massless lepton case, and only the $g_V$ term will remain. We note that the structure of of Eq.~\ref{eqn:full_amp_S} and Eq.~\ref{eqn:full_amp_P} are quite similar, except for the coupling terms and the form factors. Since $g_A$ and $g_V$ are proportional to $(1-\epsR)$ and $(1+\epsR)$, respectively, the effect of a non-zero $\epsR$ is different between the outgoing scalar and pseudoscalar meson states.

\subsubsection{The $\Bbar \to V \ellm \barnuell$ case}

When the outgoing meson is $1^-$ vector meson $V$, both the $g_V$ and $g_A$ terms contribute and the hadronic current can be written in terms of four form factors $A_0$, $A_1$, $A_2$ and $V$:
\begin{subequations}
\label{eqn:full_ffs_V}
\begin{align}
\langle V |\bar{q} \gamma_\mu b|\Bbar\rangle_V  &= \frac{2 i V(\qsq)}{m_V +m_B} \epsilon_{\mu \nu \alpha \beta} \varepsilon^{\ast \nu}_V p^\alpha_V p^\beta_B \\
\langle V |\bar{q} \gamma_\mu \gamma_5 b|\Bbar\rangle_A  &= 2 m_V A_0(\qsq) \frac{\varepsilon^\ast_V \cdot q}{q^2} q_\mu \nonumber \\ & +(m_B+ m_V) A_1(\qsq) \left(\varepsilon^\ast_{\mu V} - \frac{\varepsilon^\ast_V \cdot q}{q^2} q_\mu\right)  \nonumber \\
&- A_2(\qsq) \frac{\varepsilon^\ast_V \cdot q}{m_B + m_V}  \Big( (p_B + p_V)_\mu  \nonumber \\
& \qquad \hspace{1.2cm}- \frac{(p_B + p_V)\cdot q}{q^2} q_\mu \Big) \\
\langle V |\bar{q} \gamma_5 b|\Bbar\rangle_P  &\approx 2 m_V A_0 (\qsq) \frac{\varepsilon^\ast_V \cdot q}{m_b + m_q} \\
\langle V |\bar{q} b|\Bbar\rangle_S  &= 0,
\end{align}
\end{subequations}
and the matrix element is:
\begin{align}
\label{eqn:full_amp_V}
\mathcal{M}_V &= \frac{2 \GF V_{qb}}{\sqrt{2}} \left\{ \left[ g_V \left(\frac{2 i V(\qsq)}{m_V +m_B}  \epsilon_{\mu \nu \alpha \beta} \varepsilon^{\ast \nu}_V p^\alpha_V p^\beta_B \right) \hspace{3cm}\right. \right. \nonumber \\
&\qquad \hspace{0.2cm} - g^A \Big( (m_B+ m_V) A_1(\qsq) \varepsilon^\ast_{V \mu} - \nonumber \\
&\qquad \hspace{0.2cm} \left. A_2(\qsq) \frac{\varepsilon^\ast_V \cdot q }{m_B + M_V} (p_B + p_V)_\mu \Big) \right] \bar{\ell}\gamma^\mu \nu_L \nonumber \\
&\qquad \hspace{0.2cm} - \; g^P \left.\left(\frac{2 m_B}{m_b + m_q} \varepsilon^\ast_V \cdot q \; A_0(\qsq) \right) \bar{\ell} \nu_L \right\}.
\end{align}

\subsection{EWP decays}

The effective Hamiltonian for $b \to s$ transitions can be expanded in the form~\cite{becirevi_schneider,altman_jhep09,kruger_matias}
\begin{align}
\mathcal{H_{\mbox{\scriptsize eff} }} &=  - \frac{4 \GF}{\sqrt{2}} \displaystyle \sum_q \sum_i  V_{qb} V^\ast_{qs} (C_i \mathcal{O}_i + C'_i \mathcal{O}'_i),
\end{align}
where $\q\in\{u,c,t\}$ represents the quarks running in the loop (dominated by the heavy top quark) and $i \in \{1,...,10,S,P\}$. The unprimed and primed components represent the LH and RH (absent in the SM) hadronic currents, respectively. The $C_i$'s are the scale dependent Wilson coefficients that encode the short-distance physics, while the $\mathcal{O}_i$'s are local operators representing the non-perturbative long-distance physics. The explicit forms of the operators can be found in Ref.~\cite{becirevi_schneider}. $\mathcal{O}_{i\leq 6}$ are the 4-quark operators, suppressed at leading-order in the SM, but can contribute via charm-loop effects, especially near the charm-threshold in $\qsq$. Out of these, $\mathcal{O}_{i\leq 2}$ are tree-level operators, while $\mathcal{O}_{3\leq i \leq 6}$ are gluonic penguins. $\mathcal{O}_8$ is a gluonic dipole operator and is also suppressed by a power of $\sqrt{\alpha_s}$. The scalar and pseudoscalar operators, $\mathcal{O}_{S,P}$, do not contribute in the SM.

The three main contributing terms for the $b \to s \ellm \ellp$ EWP decays are $\mathcal{O}_{7}$,  $\mathcal{O}_{9}$ and  $\mathcal{O}_{10}$. $\mathcal{O}_{7}$ is the $\gamma^\ast$ penguin, while $\mathcal{O}_{9}$ and  $\mathcal{O}_{10}$ get contributions from the $Z^\ast$ penguin and $W^\ast$ box-diagram.

The Wilson coefficients are calculated by matching the effective and full theory at the $\mu\sim m_W$ scale, and evolved down to $\mu\sim m_b$ by the renormalization group equations. In the SM, the rough hierarchy is $C_7 \sim - 0.331$, $C_9 \sim 4.27$ ad $C_{10} \sim -4.173$, so that $C_9$ and $C_{10}$ contributions dominate, except at $\qsq \to 0$, where the photonic penguin dominates.

Following Ref.~\cite{melikhov98}, we next define the following coefficients
\begin{subequations} 
\begin{align}
\overline{C}^{L,R} &\equiv \left[ (C^{\rm eff}_9 - {C^{\rm eff}}'_9) \mp (C^{\rm eff}_{10} - {C^{\rm eff}}'_{10}) \right] /2 \\
\overline{C}'^{L,R} &\equiv \left[ (C^{\rm eff}_9 + {C^{\rm eff}}'_9) \mp (C_{10} + {C^{\rm eff}}'_{10}) \right] /2 \\
\overline{C}_{7} &\equiv (C^{\rm eff}_7 - {C^{\rm eff}}'_7)/2 \\
\overline{C}'_{7} &\equiv (C^{\rm eff}_7 + {C^{\rm eff}}'_7)/2,
\end{align}
\end{subequations} 
where $\{L,R\}$ signifies the handedness on the leptonic side and the expressions of ${C^{\rm eff}}^{(')}$ can be found in Ref.~\cite{altman_jhep09}. It should be noted that the effects of charm loops (from $C_{i\leq6}$) enter $C^{\rm eff}_9$, thereby incorporating strong phases into $C^{\rm eff}_9$.

For $X$ being in a spin-$J$ state, the helicity amplitudes in terms of the $\qsq$ dependent form-factors $F_{1,T}$, $A_{1,2}$, $V$ and $T_{1,2}$ are~\cite{luwang}:
\begin{widetext}
\begin{subequations}
\label{eqn:ewp_hel_amps_spinJ}
\begin{align}
H^{L,R} \Big|_{J=0} &=   \frac{2 m_B {\bf k}  }{\sqrt{\qsq}}\left\{  \overline{C}^{L,R} F_1(\qsq) + \overline{C}_{7} \frac{2 m_B}{m_B+m_X} F_T(\qsq) \right\} \\
H^{L,R}_\pm \Big|_{J\geq1} &= \beta_J  \left( \frac{{\bf k}}{m_X} \right)^{J-1} \Bigg\{ \Bigg. \overline{C}^{L,R} (m_B+m_X) A_1(\qsq) + \frac{2 m_B}{\qsq} (m^2_B-m^2_X) \overline{C}_{7} T_2(\qsq)  \Bigg. \nonumber\\
& \hspace{5cm} \Bigg. \mp 2 m_B {\bf k} \left[  \overline{C}'^{L,R} \frac{V(\qsq)}{m_B+m_X} + \overline{C}'_{7} \frac{2 m_B}{\qsq} T_1(\qsq) \right] \Bigg\} \\
H^{L,R}_0\Big|_{J\geq1} &= \frac{\alpha_J}{2 m_X \sqrt{\qsq}}  \left( \frac{{\bf k}}{m_X} \right)^{J-1} \Bigg\{ \overline{C}^{L,R} \left[ (m^2_B - m^2_X - \qsq)(m_B+m_X) A_1(\qsq) - \frac{4 m^2_B {\bf k}^2}{m_B+m_X}A_2(\qsq) \right] \nonumber \\
&  \hspace{5cm} \Bigg.+ 2 m_B \overline{C}_7 \left[ (m^2_B + 3 m^2_X - \qsq) T_2(\qsq) - \frac{4 m^2_B {\bf k}}{m^2_B - m^2_X} T_3(\qsq) \right] \Bigg\}.
\end{align}
\end{subequations}
\end{widetext}
The $\{\alpha_J,\beta_J\}$ factors come from Clebsch-Gordon coefficients and are $\{1,1\}$ and $\{\sqrt{2/3},1/\sqrt{2}\}$ for the vector and tensor states, respectively~\cite{luwang}. The ${\bf k}^{J-1}$ terms are additional kinematic factors incorporating the angular momentum barrier factors for higher spins (see also discusion in Sec.~\ref{sec:mass_dep}).

Note that in the above equations for the helicity amplitudes, relative to the convention in the EWP literature, we have taken out an overall normalization factor. The benefit is that the SL limit is easily arrived at by the substititions $\overline{C}^{r} = \overline{C}'^r = \overline{C}_{7} = \overline{C}'_{7} =0$, and $\overline{C}^l = \overline{C}'^l=1$ (see also Eq.~\ref{eqn:sl_ewp}). The terms corresponding to $F_T$ and $T_i$'s do not exist in the SL case and $F_1$ is identified as $u_+(f_+)$ for $\bar{B} \to S(P) \ellm \bar{\nu}_\ell$.

\section{Differential rate for $\Bbar \to X \ell_1 \ell_2$}

Following Hagiwara~\cite{hagiwara_plb,hagiwara_npb}, the differential rate is
\begin{equation}
\label{eqn:diff_rate_btoxlnu}
d\Gamma = \frac{1}{2 m_B} \displaystyle \sum_{\mbox{\scriptsize final spins}} | \mathcal{M} |^2 d \phi_3,
\end{equation}
where the incoherent sum is over the spins of all final-state particles and three-body $X \ell_1 \ell_2$ phase-space factor is
\begin{equation}
\label{eqn:three_body_ps_xlnu}
d \phi_3 = \frac{{\bf k}}{2 m_B} \frac{d\qsq d \ctl}{(4 \pi)^3}.
\end{equation}
where ${\bf k}$ is the usual $X$ 3-momentum magnitude in the $B$ RF. Including only spin-0 and spin-1 states for the di-lepton system, the invariant amplitude can be written in the form
\begin{equation}
\mathcal{M} = \frac{G_F V}{\sqrt{2}}\left\{ (\mathcal{H}_P + \mathcal{H}_S) L_S + \sum_{\eta = \pm 1} \sum_{\lambda \in \{0,\pm 1\}} L^\eta_{\lambda} \mathcal{H}^\eta_\lambda \right\},
\end{equation}
where $\lambda$ is the helicity of the hadronic system $X$ and the handedness $\eta \equiv (\lambda_{\ell_1} - \lambda_{\ell_2}) = -1(+1)$ for LH(RH) leptonic currents. For SL decays with $\ell_1$ always being the charged lepton, for a $(\ellm \barnuell)$ final state, since $\lambda = +1/2$ for the purely RH $\barnuell$, we have $\eta = -1$. For the $(\ell^+ \nu)$ final state, we have $\eta = +1$ for the purely LH $\nu$. Here, $V$ is a scale factor that equals $G_FV_{qb}$ in SL $b\to q$ type transitions. For the EWP decays $\Bbar \to \overline{K}^{(\ast)} \ellm \ellp$, the effective replacement is~\cite{melikhov98}:
\begin{align}
\label{eqn:sl_ewp}
V_{qb}\Big|_{\rm SL} \to \displaystyle \left(\frac{\alpha}{2 \pi} V^\ast_{ts} V_{tb} \right) \Big|_{\rm EWP}.
\end{align}

The hadronic helicity amplitudes for $\Bbar$ (that is, containing a $b$ quark) are defined as~\cite{korner_schilcher}
\begin{equation}
\mathcal{H}^\eta_\lambda = \left(\varepsilon^\ast_{W^\ast}(\lambda)\right)_{\mu} \langle X(\lambda)|J^\mu|\Bbar\rangle_\eta,
\end{equation}
with the spin-quantization axis along the $X$ flight direction in the $\Bbar$ RF, while the leptonic helicity amplitudes are (for massless leptons)
\begin{equation}
L^\eta_\lambda = 2 \left(\varepsilon_{W^\ast}(-\lambda)\right)_{\mu} \bar{u}_{\ell_1} \gamma^\mu  u_{\ell_2} = 2 \sqrt{2\qsq} d^1_{\lambda,\eta} (\thetal)
\end{equation}
where the spin-quantization axis is along the di-lepton flight direction in the $\Bbar$ RF, opposite to the $X$ direction. Since the parent meson is spin-0, the helicities of the daughter hadronic and the leptonic systems have to be the same.

For the scalar term $L_S$, the helicities of the two leptons must be the same, since the total spin of the di-lepton system is 0. This means, that although~\cite{hagiwara_plb,hagiwara_npb} $\bar{\ell} \nu_L = \sqrt{\qsq}$, so that
\begin{equation}
L_S = 2 \bar{\ell} \nu_L = 2 \sqrt{\qsq},
\end{equation}
the lepton spin-configurations for the $L_\lambda$ and $L_S$ cases are different and the two components must add incoherently in total rate expression. The spin-0 leptonic current terms for the massless lepton case are therefore second order corrections relative to the SM, and will be neglected henceforth.

\subsection{The $\Bbar \to P(S) \ellm \barnuell$ case}

Following the calculations in Ref.~\cite{hagiwara_npb}, one can show that the amplitude in Eq.~\ref{eqn:full_amp_P} for the SL outgoing pseudoscalar meson case is
\begin{equation}
\mathcal{M}_P = \frac{\GF V_{qb}}{\sqrt{2}}g_V (-4 m_B {\bf k} \sin \thetal f_+(\qsq)),
\end{equation}
where we have neglected the $g_S$ term because, as explained above, it is a small second order correction to the SM. Plugging this into Eqs.~\ref{eqn:diff_rate_btoxlnu} and~\ref{eqn:three_body_ps_xlnu}, we get
\begin{equation}
\label{eqn:full_angular_distribution_P}
\frac{d\Gamma}{d\phi} = \frac{\GF^2 |V_{qb}|^2}{32\pi^3}  {\bf k}^3 \sin^2\thetal |g_V f_+(\qsq)|^2,
\end{equation}
where $d\phi = d\qsq d\cos\thetal$.

The outgoing scalar meson case is obtained by replacing $g_V$ with $ g_A$, and $f_+$ with $u_+$.

\subsection{The $X \to P_1 P_2$ case}

For the case where the $X$ system decays into a pair of spin-0 pseudo-scalars the amplitude can be expanded as
\begin{equation}
\label{eqn:hel_amp}
\mathcal{M} = \frac{G_F V}{\sqrt{2}} \left\{\displaystyle \sum_{\eta = \pm 1}  \sum_{\lambda \in \{0,\pm1\}}  L^\eta_\lambda  \mathcal{H}^\eta_\lambda Y_\lambda\right\}
\end{equation}
where, for the hadronic system, 
\begin{align}
\label{eqn:hel_amp_exp}
\mathcal{H}^\eta_\lambda Y_\lambda \equiv \sum_J \mathcal{H}^{{\eta,J}}_\lambda Y^\lambda_J(\thetav,\chi).
\end{align}
The spherical harmonics are given in terms of the Wigner $d$-matrices as
\begin{align}
\label{eqn:ylm}
Y^\lambda_J(\thetav,\chi) \equiv \displaystyle \sqrt{\frac{2J+1}{4 \pi}} d^J_{\lambda,0}(\thetav)e^{i\lambda \chi}
\end{align}
and the differential phase-space element is now ${d\phi = d\phi_3 d\cos \thetaV d \chi}$

Putting everything together, the full expression for the 4-D differential rate for $X$ decaying to two pseudoscalars is then
\begin{align}
\label{eqn:rate_dstodpi}
\frac{d\Gamma}{d\phi} &= \frac{ |V|^2 {\bf k} \qsq \mathcal{B}^{X\to P_1 P_2} }{m^2_B (4 \pi)^4} |\overline{\mathcal{M}}|^2
\end{align}
where,
\begin{align}
\label{eqn:rate_M2_XP1P2}
|\overline{\mathcal{M}}|^2 &= \sum_{\eta = \pm 1}  \Bigg| \sum_{\lambda\in \{0,\pm 1\}} \sum_J \sqrt{2J+1} \mathcal{H}^{\eta,J}_\lambda d^J_{\lambda,0} (\thetaV) \nonumber \\
& \hspace{5cm} d^1_{\lambda,\eta} (\thetal) e^{i\lambda \chi} \Bigg|^2,
\end{align}
and $\mathcal{B}^{X\to P_1 P_2} $ is the relevant branching fraction (BF). The LH and RH contributions add incoherently since the final-state spin configurations are different on the leptonic side. 

For $J\in\{0,1,2\}$, we denote the spin-0, spin-1 and spin-2 helicity amplitudes as $S^{\{L,R\}}$, $H^{\{L,R\}}_\lambda$ and $D^{\{L,R\}}_\lambda$, respectively, where the superscripts denote the handedness of the leptonic current. The full expansion of $|\overline{\mathcal{M}}|^2$ yields 41 angular terms:
\begin{align}
|\overline{\mathcal{M}}|^2 = \frac{1}{16} \displaystyle \sum_{i=1}^{41} (h^L_i + \eta^{L\to R}_i h^R_i),
\label{eqn:spd_expansion_hel}
\end{align}
as tabulated in Table~\ref{table:spd_expansion_hel}. Here, $\eta^{L\to R}_i=\pm 1$ is a sign factor dictated by the behavior of the angular part under $\thetal \to \pi + \thetal$, since $d^1_{\lambda,\eta}(\thetal) \equiv d^1_{\lambda,-\eta}(\pi +\thetal)$. $h^R_i$ is of the same form as $h^L_i$, except with all the LH amplitudes replaced by their RH counterparts.

\begin{table*}
\begin{center}
\renewcommand{\arraystretch}{1.31}
\begin{tabular}{c|c|c} 
 $i$ &   $h^L_i(\phi)$  & $\eta^{L\to R}_i$ \\ \hline \hline
  1& $6(|H^L_+|^2 + |H^L_-|^2) + 8 |S^L|^2 + 10 |D^L_0|^2 - 8 \sqrt{5} \rel(S^L D^{L\ast}_0) $ & +($L\to R)$ \\ \hline
  2& $\ctv [12 \sqrt{5} \rel(H^L_+ D^{L\ast}_+ + H^L_- D^{L\ast}_-) + 16 \sqrt{3} \rel(S^L H^{L\ast}_0) - 8 \sqrt{15} \rel(D^L_0 H^{L \ast}_0)] $ & " \\ \hline
  3& $\cos^2 \thetav [30 (|D^L_+|^2 + |D^L_-|^2) - 6(|H^L_+|^2 + |H^L_-|^2) + 24|H^L_0|^2 + 24\sqrt{5}\rel(S^L D^{L\ast}_0) - 60|D^L_0|^2] $ & " \\ \hline
  4& $\cos^3 \thetav [ -12 \sqrt{5} \rel(H^L_+ D^{L\ast}_+ + H^L_- D^{L\ast}_-) + 24 \sqrt{15} \rel(H_0 D^{L\ast}_0)] $ & " \\ \hline
  5& $\cos^4 \thetav [-30 (|D^L_+|^2 + |D^L_-|^2) + 90 |D^L_0|^2)] $ & " \\ \hline
  6& $\cos^2\thetal[6(|H^L_+|^2 + |H^L_-|^2) - 8 |S^L|^2 - 10 |D^L_0|^2 + 8 \sqrt{5} \rel(S^L D^{L\ast}_0)] $ & " \\ \hline
  7& $\cos^2\thetal \ctv   [12 \sqrt{5} \rel(H^L_+ D^{L\ast}_+ + H^L_- D^{L\ast}_-) - 16 \sqrt{3} \rel(S H^\ast_0) + 8 \sqrt{15} \rel(D_0 H^\ast_0)] $ & " \\ \hline
  8& $\cos^2\thetal \cos^2 \thetav [30 (|D^L_+|^2 + |D^L_-|^2) - 6(|H^L_+|^2 + |H^L_-|^2) - 24|H^L_0|^2 - 24 \sqrt{5} \rel(S^L D^{L\ast}_0) + 60|D^L_0|^2] $ & " \\ \hline
  9& $\cos^2\thetal \cos^3 \thetav  [ -12 \sqrt{5} \rel(H^L_+ D^{L\ast}_+ + H^L_- D^{L\ast}_-) - 24 \sqrt{15} \rel(H^L_0 D^{L\ast}_0)] $ & " \\ \hline
  10& $\cos^2\thetal \cos^4 \thetav [-30 (|D^L_+|^2 + |D^L_-|^2) - 90 |D^L_0|^2)] $ & " \\ \hline
  11& $\stl \ctl \cos \chi \stv [-8 \sqrt{3} \rel((H^L_+ + H^L_-)S^\ast) + 4\sqrt{15} \rel((H^L_+ + H^L_-)D^{L\ast}_0) ] $ & " \\ \hline
  12& $\stl \ctl \cos \chi \stv \ctv [-24\rel((H^L_+ + H^L_-) H^{L\ast}_0) - 8\sqrt{15} \rel((D^L_+ + D^L_-) S^{L\ast}) + 20 \sqrt{3} \rel((D^L_+ + D^L_-)D^{L\ast}_0  )   ] $ & " \\ \hline
  13& $\stl \ctl \cos \chi \stv \cos^2 \thetav[-24\sqrt{5} \rel((D^L_+ + D^L_-) H^{L\ast}_0  - 12 \sqrt{15} \rel((H^L_+ + H^L_-) D^{L\ast}_0)] $ & " \\ \hline
  14& $\stl \ctl \cos \chi \stv \cos^3 \thetav[-60 \sqrt{3} \rel((D^L_+ + D^L_-) D^{L\ast}_0 )] $ & " \\ \hline
  15& $\stl \ctl \sin \chi \stv[ 8 \sqrt{3}\img( (H^L_+ - H^L_-)S^{L\ast}) - 4\sqrt{15} \img( (H^L_+ - H^L_-)D^{L\ast}_0)] $ & " \\ \hline
  16& $\stl \ctl \sin \chi \stv \ctv[ 24 \img((H^L_+ - H^L_-) H^{L\ast}_0) + 8\sqrt{15}\img((D^L_+ - D^L_-)S^{L\ast}) -  20 \sqrt{3}\img((D^L_+ - D^L_-) D^{L\ast}_0)] $ & " \\ \hline
  17& $\stl \ctl \sin \chi \stv \cos^2 \thetav[ 24 \sqrt{5} \img((D^L_+ - D^L_-) H^{L\ast}_0) + 12 \sqrt{15} \img((H^L_+ - H^L_-) D^{L\ast}_0)] $ & " \\ \hline
  18& $\stl \ctl \sin \chi \stv \cos^3 \thetav[ 60\sqrt{3}  \img((D^L_+ - D^L_-) D^{L\ast}_0)] $ & " \\ \hline
  19& $\sin^2 \thetal \cos 2 \chi[ -12 \rel( H^L_+ H^{L\ast}_-)] $ & " \\ \hline
  20& $\sin^2 \thetal \cos 2 \chi \ctv[ -12 \sqrt{5} \rel(H^L_+ D^{L\ast}_- + D^L_+ H^{L\ast}_-)] $ & " \\ \hline
  21& $\sin^2 \thetal \cos 2 \chi \cos^2 \thetav[ -60\rel(D^L_+ D^{L\ast}_-) + 12 \rel(H^L_+ H^{L\ast}_-)] $ & " \\ \hline
  22& $\sin^2 \thetal \cos 2 \chi \cos^3 \thetav[ 12 \sqrt{5} \rel(H^L_+ D^{L\ast}_- + D^L_+ H^{L\ast}_-)] $ & " \\ \hline
  23& $\sin^2 \thetal \cos 2 \chi \cos^4 \thetav[ 60 \rel(D^L_+ D^{L\ast}_- )] $ & " \\ \hline
  24& $\sin^2 \thetal \sin 2 \chi[ 12 \img(H^L_+ H^{L\ast}_- )] $ & " \\ \hline
  25& $ \sin^2 \thetal \sin 2 \chi \ctv [ 12 \sqrt{5} \img(H^L_+ D^{L\ast}_- + D^L_+ H^{L\ast}_-)  ] $ & " \\ \hline
  26& $\sin^2 \thetal \sin 2 \chi \cos^2 \thetav[ 60 \img(D^L_+ D^{L\ast}_-) - 12 \img(H^L_+ H^{L\ast}_-)] $ & " \\ \hline
  27& $\sin^2 \thetal \sin 2 \chi \cos^3 \thetav[ -12 \sqrt{5}  \img(H^L_+ D^{L\ast}_- + D^L_+ H^{L\ast}_-)] $ & " \\ \hline
  28& $\sin^2 \thetal \sin 2 \chi \cos^4 \thetav[ -60 \img(D^L_+ D^{L\ast}_-)] $ & " \\ \hline \hline
  29& $\ctl [12 (|H^L_-|^2 - |H^L_+|^2)] $ & -($L\to R$) \\ \hline
  30& $\ctl \ctv[ 24 \sqrt{5} \rel(H^L_- D^{L\ast}_- - H^L_+ D^{L\ast}_+) ] $ & " \\ \hline
  31& $\ctl \cos^2 \thetav[ 60 (|D^L_-|^2 - |D^L_+|^2) - 12 (|H^L_-|^2 - |H^L_+|^2) ] $ & " \\ \hline
  32& $\ctl \cos^3 \thetav[-24 \sqrt{5} \rel(H^L_- D^{L\ast}_- - H^L_+ D^{L\ast}_+) ] $ & " \\ \hline
  33& $\ctl \cos^4 \thetav[- 60 (|D^L_-|^2 - |D^L_+|^2)] $ & " \\ \hline
  34& $\stl \cos \chi \stv[ 8 \sqrt{3} \rel((H^L_+ - H^L_-)S^\ast) - 4\sqrt{15} \rel( (H^L_+ - H^L_-)D^{L\ast}_0)] $ & " \\ \hline
  35& $\stl \cos \chi \stv \ctv[ 24 \rel((H^L_+ - H^L_-) H^{L\ast}_0) + 8\sqrt{15} \rel((D^L_+ - D^L_-)S^{L\ast}) - 20 \sqrt{3}\rel((D^L_+ - D^L_-) D^{L\ast}_0)] $ & " \\ \hline
  36& $\stl \cos \chi \stv \cos^2 \thetav[ 24 \sqrt{5} \rel((D^L_+ - D^L_- )H^{L\ast}_0) + 12 \sqrt{15} \rel((H^L_+ - H^L_-) D^{L\ast}_0)] $ & " \\ \hline
  37& $\stl \cos \chi \stv \cos^3 \thetav[ 60 \sqrt{3} \rel((D^L_+ - D^L_-) D^{L\ast}_0) ] $ & " \\ \hline
  38& $ \stl \sin \chi \stv[ -8 \sqrt{3} \img((H^L_+ + H^L_-)S^{L\ast}) + 4\sqrt{15} \img((H^L_+ + H^L_-)D^{L\ast}_0)] $ & " \\ \hline
  39& $\stl \sin \chi \stv \ctv[ -24 \img((H^L_+ + H^L_-) H^{L\ast}_0) - 8 \sqrt{15}\img((D^L_+ + D^L_-)S^{L\ast}) + 20 \sqrt{3} \img((D^L_+ + D^L_-)D^{L\ast}_0)] $ & " \\ \hline
  40& $\stl \sin \chi \stv \cos^2 \thetav[ -24 \sqrt{5} \img((D^L_+ + D^L_-) H^{L\ast}_0) - 12 \sqrt{15} \img((H^L_+ + H^L_-) D^{L\ast}_0)] $ & " \\ \hline
  41& $ \stl \sin \chi \stv \cos^3 \thetav[ -60 \sqrt{3} \img((D^L_+ + D^L_-) D^{L\ast}_0 )] $ & " \\ \hline 
\end{tabular}
\caption{The helicity-basis expansion of $|\overline{\mathcal{M}}|^2$ in Eq.~\ref{eqn:spd_expansion_hel}.}
\label{table:spd_expansion_hel}
\end{center}
\end{table*}

\subsection{The $\Bbar \to V  \ellm \barnuell$ case}

The $V \to P_1 P_2$ case is the same as Eq.~\ref{eqn:rate_dstodpi}, with only the $\mathcal{H}^L_\lambda$ amplitudes contributing. For $V \to P \gamma$, we need to incoherently sum over the outgoing photon helicity $\lambda_\gamma = \pm1$ cases separately:
\begin{align}
\label{eqn:rate_dstodgamma}
\frac{d\Gamma}{d\phi} &= \frac{ 3 \GF^2 |V_{qb}|^2 {\bf k} \qsq \mathcal{B}^{V \to P \gamma} }{m^2_B (4 \pi)^4} \nonumber \\
& \qquad \displaystyle \sum_{\lambda_\gamma = \pm 1} \frac{1}{2}\left| \sum_{\lambda\in \{0,\pm 1\}} \mathcal{H}^L_\lambda d^1_{\lambda,\lambda_\gamma} (\thetaV) d^1_{\lambda,-1} (\thetal) e^{i\lambda \chi} \right|^2 \nonumber \\
&= \frac{ 3 \GF |V_{qb}|^2 {\bf k} \qsq \mathcal{B}^{V \to P \gamma} }{32 m^2_B (4 \pi)^4} \nonumber \\
&\qquad \times \displaystyle \sum_{\lambda_\gamma = \pm 1} \Big|2 \lambda_\gamma \sin\thetaV \mathcal{H}^L_0 (-\sin\thetal) \nonumber \\ 
&\qquad \hspace{1cm} + \mathcal{H}^L_{+1} (1 + \lambda_\gamma \cos\thetaV)(1- \ctl)e^{i\chi}  \nonumber \\
&\qquad  \hspace{1cm} + \mathcal{H}^L_{-1} (1 - \lambda_\gamma \cos\thetaV)(1+ \ctl)e^{-i\chi} \Big|^2
\end{align}
where the extra factor of $\frac{1}{2}$ ensures normalization to the appropriate BFs.

We write $\mathcal{H}^L_\lambda \equiv H_\lambda e^{i \delta_\lambda}$, and set $\delta_0 = 0$. For $\Bbar \to V  \ellm \barnuell$, the expressions in Eqs.~\ref{eqn:rate_dstodpi} and~\ref{eqn:rate_dstodgamma} can then be summarized as:
\begin{widetext}
\begin{align}
\label{eqn:full_angular_distribution_V}
\frac{d\Gamma }{d\phi} &= \left[ \frac{\mathcal{C}'}{ 1+(1-\alpha)/2} \right] \times \nonumber \\ 
&\qquad \Bigg\{\Big[ H_+^2  (1 -\cos\thetal)^2 +   H_-^2 (1 +\cos\thetal)^2\Big](1-\alpha \cos^2\thetaV) + 4 H_0^2 \sin^2\thetal \Big(\frac{1-\alpha}{2}+ \alpha\cos^2\thetaV \Big) \Bigg.\nonumber  \\ 
&\qquad + 2 \alpha H_0 \sin\thetal \sin2\thetaV \Big[ H_+ (1-\cos\thetal) \cos (\chi + \delta_+) - H_- (1+\cos\thetal) \cos(\chi - \delta_-)\Big] \nonumber \\ 
&\qquad \Bigg.- 2\alpha  H_+H_- \sin^2\thetal \sin^2\thetaV\cos(2 \chi + (\delta_+ - \delta_-)) \Bigg\},
\end{align}
\end{widetext}
where $\alpha$ is -1 for $V\to P\gamma$ (such as $D^\ast \to D \gamma$ or $\omega \to \pi \gamma$) and +1 for $V\to P_1 P_2$ (such as $\rho \to \pi \pi$ or $D^\ast \to D \pi$) type decays. The pre-factor term is
\begin{equation}
\label{eqn:pre_factor_SL}
\mathcal{C}' = \frac{3}{8(4\pi)^4} |V_{qb}|^2 \frac{{\bf k}  \qsq}{m^2_B} \BR
\end{equation}
where the term $\BR$ accomodates any BFs from the vector meson decay chain onwards. 
 
For the $V\to P_1 P_2$ type cases, Eq.~\ref{eqn:full_angular_distribution_V} above agrees with Eq.~2.20 in Ref.~\cite{gilman_full_expr}. It also agrees with Eq.~113 in Ref.~\cite{richman_burchat_SL} after taking into account the change in the $\chi$ definition as given by Eq.~\ref{eqn:gs2rb}c.

\section{Incorporating mass-dependences}
\label{sec:mass_dep}

When the variation of the invariant mass of the $X$ system, $m \equiv m_X$, is no longer negligible, Eq.~\ref{eqn:rate_dstodpi} can be extended as
\begin{align}
\label{eqn:rate_dstodpi_mass_dep}
\frac{d\Gamma}{d\phi dm} &= \left(\frac{p}{p_0} \right) \frac{ |V|^2 {\bf k} \qsq \mathcal{B}^{X\to P_1 P_2}(m) }{m^2_B (4 \pi)^4} |\overline{\mathcal{M}}(m)|^2,
\end{align}
where $p$ is the mass-dependent breakup momentum of $X\to P_1 P_2$ in the $X$ rest frame, $p_0$ is the value of $p$ computed at the (dominant) pole mass $m_0$. The overall factor of $p/p_0$ comes from phase-space.

The $\mathcal{H}^J$ amplitudes incorporate a mass-dependent relativistic Breit-Wigner (rBW) part
\begin{align}
\mathcal{H}^J_{\rm rBW}(m) \sim \left(\frac{p}{p_0}\right)^J \frac{B^J(p,R)}{B^J(p_0,R)} \frac{1}{m^2_0 - m^2 - im_0\Gamma^J_{\mbox{\scriptsize total}}},
\end{align}
where $B^J(p,R)$ is the phenomenological Blatt-Weisskopf barrier factor with $R \approx \mathcal{O}(5~\rm{GeV}^{-1})$, corresponding to a meson radius of $\mathcal{O}(1~\rm{fm})$. For a $P$-wave decay this is given by~\cite{pdg}
\begin{align} 
B^{J=1}(p,R) = \frac{1}{\sqrt{1+p^2R^2}}.
\end{align}
For the spin-$J$ resonance having a single decay mode to the final state $P_1 P_2$, 
\begin{align}
\Gamma^J_{\mbox{\scriptsize total}} = \Gamma_0 \left(\frac{p/m}{p_0/m_0}\right) \left| \left(\frac{p}{p_0}\right)^J \frac{B^J(p,R)}{B^J(p_0,R)}\right|^2.
\end{align}
However, if the spin-$J$ resonance has $k$ decay modes, all the individual mass-dependent widths contribute as
\begin{align}
\Gamma^J_{\mbox{\scriptsize total}} = \displaystyle \sum_{i=1}^{k} \Gamma^J_i \mathcal{B}^J_i,
\end{align}
where $\mathcal{B}^J_i$ is the branching fraction into the $i^{\rm th}$ mode. Examples of such instances are the decay modes of the $\phi(1020)$ or the $K^\ast_2(1430)$.

The second form of mass-dependence comes from the barrier factor associated with the $B$ decay itself. Let the $B$ decay into the di-lepton and $X$ system occur with an angular momentum $L_B$ and the break-up momentum is ${\bf k}$, as given by Eq.~\ref{eqn:k_expr}. If the $X$ system is in spin-$J$, the selection rule is $L_B\in\{J-1,J,J+1\}$. The helicity amplitudes $\mathcal{H}^J_\lambda$ can be re-written in terms of specific $L_B$ components with the relevant Clebsch-Gordon factor $\langle J,\lambda;1,-\lambda|L_B,0\rangle$ as:
\begin{subequations} 
\begin{alignat}{5}
&S   \;& \equiv& \; S^1 &&\\
&H_\pm \;& \equiv& \; \frac{1}{\sqrt{6}} H^2 &\pm \frac{1}{\sqrt{2}} H^1 &+ \frac{1}{\sqrt{3}} H^0\\
&H_0 \;& \equiv& \; \sqrt{\frac{2}{3}}H^2 &&- \frac{1}{\sqrt{3}} H^0 \\
&D_\pm \;& \equiv& \; \frac{1}{\sqrt{5}} D^3 &\pm  \frac{1}{\sqrt{2}} D^2 &+ \sqrt{\frac{3}{10}}  D^1 \\
&D_0 \;& \equiv& \; \sqrt{\frac{3}{5}} D^3 && - \sqrt{\frac{2}{5}} D^1
\end{alignat}
\end{subequations} 
The superscripts on the rhs denote the $L_B$ values, and the amplitudes represent the spin-$L_B$ component of the corresponding helicity amplitude. 
Each spin-$L_B$ component of the helicity amplitudes acquires a nominal barrier factor that scales as $b = {\bf k}^{L_B} B^{L_B}({\bf k},R)$. We define the normalized quantity $x_{L_B} =  b/b_0$, where we choose to calculate the denominator at the pole mass. The mass-dependent helicity amplitudes are

\begin{subequations} 
\label{eqn:mass_dep_hel_amps}
\begin{alignat}{3}
&S(m) \;&\equiv&\; x_1 S\\
&H_\pm(m) \;&\equiv&\; \frac{x_2}{3} \left( H_0 + \frac{H_+ + H_-}{2}\right) + \frac{H_+ + H_- - H_0}{3} \nonumber \\
&& & \hspace{1cm} \pm x_1 \left(\frac{H_+ - H_-}{2}\right) \\
&H_0(m)   \;&\equiv&\; \frac{2 x_2}{3} \left( H_0 + \frac{H_+ + H_-}{2} \right) - \frac{H_+ + H_- - H_0}{3} \\
& D_\pm(m)   \;&\equiv&\;  \frac{x_3}{5} \left( \sqrt{3} D_0 + (D_+ + D_-)\right) \nonumber \\
&& & \hspace{1cm}  + x_1 \sqrt{\frac{3}{10}} \left(\sqrt{\frac{3}{10}} (D_+ + D_-) - \sqrt{\frac{2}{5}} D_0\right) \nonumber \\
&& & \hspace{1cm} \pm x_2 \left( \frac{D_+ - D_-}{2}  \right)  \\      
& D_0(m)   \;&\equiv&\;  \frac{x_3 \sqrt{3}}{5} \left( \sqrt{3} D_0 + (D_+ + D_-)\right) \nonumber \\
&& & \hspace{1cm}  - x_1 \sqrt{\frac{2}{5}} \left(\sqrt{\frac{3}{10}} (D_+ + D_-) - \sqrt{\frac{2}{5}} D_0\right).
\end{alignat}
\end{subequations} 
The mass-independent forms are obtained by the substitutions $x_{L_B} \to 1$. For the SL and EWP cases, the ${\bf k}$-dependent barrier factors were already incorporated in Eqs.~\ref{eqn:ewp_hel_amps_spinJ}. For the $c \bar{c}\to \ellm \ellp$ decays, the mass-dependent forms in Eq.~\ref{eqn:mass_dep_hel_amps} are more appropriate than the bare amplitudes.

\section{CP conjugation}
\label{sec:cp_conj}

Consider the CP conjugation of the process $\Bbar \to X(\to P_1 Y) \ell_1 \ell_2$, where $P_1$ is a charged pseudoscalar meson, and $\ell_1$ is a charged lepton. The CP conjugate process is $B \to \bar{X}(\to \bar{P_1} \bar{Y}) \bar{\ell_1} \bar{\ell_2}$. We perform the CP conjugation explicitly. That is, for the construction of the angular variables, going from $\Bbar$ to $B$, we replace the 4-momenta as $p_X\to p_{\bar{X}}$, $p_{P_1}\to p_{\bar{P_1}}$, $p_{\ell_1} \to p_{\bar{\ell_1}}$ and $p_{\ell_2} \to p_{\bar{\ell_2}}$. This construction leads to $\chi \to -\chi$. 

On the other hand, the effect of CP conjugation on the helicity amplitudes flips the helicities and weak phases
\begin{align}
\overline{\mathcal{H}}^\eta_\lambda(\delta_W,\delta_s) = \mathcal{H}^{-\eta}_{-\lambda}(-\delta_W,\delta_s),
\label{eqn:cp_conj_amps}
\end{align}
where $\delta_W(\delta_S)$ is any weak(strong) phase and flipping the sign of $\eta$ changes the LH amplitudes to the RH amplitudes. In the absense of direct CP violation, the simultaneous effect of these two transformations is to leave $|\overline{\mathcal{M}}|^2$ unchanged in Eq.~\ref{eqn:rate_M2_XP1P2}. Therefore, with explicit CP conjugation of the particles during construction of the angular variables (see appendix for details), no additional changes to the rate equation are required.

We also stress here that our unbarred amplitudes are defined for the $\overline{B}$ (or $b$ quark) decay, in contrast to conventions in $c \bar{c}$ analyses~\cite{babar_verderi2005, babar_verderi2007}, where the unbarred (barred) amplitudes are defined for the $B$ ($\overline{B}$) decay.   

\section{Expansion in an orthonormal basis}

Equation~\ref{eqn:rate_dstodpi} can be expanded in an orthonormal basis of angular functions $f_i(\Omega)$ as follows
\begin{subequations}
\label{eqn:vector_moments}
\begin{align}
\frac{d\Gamma }{d\qsq d\Omega} &= \mathcal{C} \times \left\{ \displaystyle \sum^{41}_{i=1} f_i (\Omega) \Gamma_i(\qsq) \right\} \\
\Gamma_i(\qsq) &= \Gamma^L_i(\qsq) + \eta^{L\to R}_i\; \Gamma^R_i(\qsq),
\end{align}
\end{subequations}
where $d\Omega = d\ctl d\ctv d\chi$ and the $\Gamma_i^{\{L,R\}}$ superscripts in Eq.~\ref{eqn:vector_moments}b specify the LH or RH nature of the leptonic current. The sign $\eta^{L\to R}_i=\pm 1$ depends on the signature of $f_i$ under $\thetal \to \pi + \thetal$. Orthonormality of the $f_i$'s imply
\begin{align}
\int f_i(\Omega) f_j(\Omega) & d \Omega = \delta_{ij}.
\end{align}
The orthonomal angular basis is constructed out of the the spherical harmonics $Y^m_l \equiv Y^m_l (\thetal,\chi)$ and the reduced spherical harmonics ${P^m_l \equiv \sqrt{2 \pi}Y^m_l(\thetav,0)}$. The pre-factor is 
\begin{equation}
\mathcal{C} = \frac{ \sqrt{8 \pi}  |V|^2 {\bf k} \qsq \mathcal{B}^{X\to P_1 P_2} }{3 m^2_B (4 \pi)^4}
\end{equation}
Defining the transversity basis amplitudes $\mathcal{H}^J_{\{\parallel, \perp\}}$ as
\begin{align}
\mathcal{H}^J_\pm &= (\mathcal{H}^J_\parallel\pm \mathcal{H}^J_\perp)/\sqrt{2},
\end{align}
Tables~\ref{table:spd_mom_hel} and~\ref{table:spd_mom_trans} list the 41 moments in the helicity and transversity bases, respectively. 

We note that since the RH and LH amplitudes are equal for the $c\bar{c} \to \ellm \ellp$ type decays, the terms with $\eta^{L\to R} = -1$ vanish and only 28 non-zero moments survive in Tables~\ref{table:spd_mom_hel} and~\ref{table:spd_mom_trans} for these cases.

\begin{table*}
\begin{center}
\renewcommand{\arraystretch}{1.31}
\begin{tabular}{c|c|c|c} 
 $i$    &   $f_i(\Omega)$             & $\Gamma^{L,{\rm hel}}_i(\qsq)$ & $\eta^{L\to R}_i$ \\ \hline \hline
 1   &   $P^0_0 Y^0_0$     &  $\left[ \hzsq + \hpsq + \hmsq + \ssq + \dzsq + \dpsq + \dmsq\right]$ & + ($L \to R$)\\ \hline 
 2   &   $P^0_1 Y^0_0$     &  $2\left[\frac{2}{\sqrt{5}} \rhzdz + \rshz + \sqrt{\frac{3}{5}}  \rel(\hpdp)\right]$ & " \\ \hline 
 3   &   $P^0_2 Y^0_0$     &  $\frac{\sqrt{5}}{7}$ (\dpsq + \dmsq) - $\frac{1}{\sqrt{5}}$ (\hpsq + \hmsq) + $\frac{2}{\sqrt{5}}$ \hzsq  + $\frac{10}{7\sqrt{5}}$ \dzsq + $2$ \rsdz & " \\  \hline
 4   &   $P^0_3 Y^0_0$     &  $\frac{6}{\sqrt{35}} \left[ - \rel(H^L_+ D^{L\ast}_+ +  H^L_- D^{L\ast}_-)  + \sqrt{3} \rhzdz  \right]$ & "\\  \hline
 5   &   $P^0_4 Y^0_0$     &  $\frac{2}{7} \left[ -2 (\dpsq + \dmsq) + 3 \dzsq \right] $ & "\\  \hline
 6   &   $P^0_0 Y^0_2$     &  $\frac{1}{2 \sqrt{5}} \left[ (\dpsq + \dmsq) + (\hpsq + \hmsq) - 2 \ssq - 2 \dzsq - 2 \hzsq \right]$ & " \\  \hline
 7   &   $P^0_1 Y^0_2$     &  $\left[ \frac{\sqrt{3}}{5} \rel(H^L_+ D^{L\ast}_+  + H^L_- D^{L\ast}_-) - \frac{2}{\sqrt{5}} \rel(S^L H^{L\ast}_0)  - \frac{4}{5} \rel(H^L_0 D^{L\ast}_0)\right] $ & "  \\ \hline
 8   &   $P^0_2 Y^0_2$     &  $ \left[ \frac{1}{14} (\dpsq + \dmsq) - \frac{2}{7} \dzsq - \frac{1}{10} (\hpsq + \hmsq) - \frac{2}{5} \hzsq - \frac{2}{\sqrt{5}} \rsdz \right]$ & "  \\  \hline
 9   &   $P^0_3 Y^0_2$     &  $ - \frac{3}{5 \sqrt{7}} \left[ \rel( H^L_+ D^{L \ast}_+ + H^L_- D^{L \ast}_-) + 2 \sqrt{3} \rel(H^L_0 D^{L \ast}_0 ) \right] $ & "\\  \hline
 10  &   $P^0_4 Y^0_2$     &  $ -\frac{2}{7 \sqrt{5}}  \left[ \dpsq + \dmsq + 3 \dzsq \right] $ & "  \\  \hline
 11  &   $P^1_1 \sqrt{2}\rel(Y^1_2)$ &  $-\frac{3}{\sqrt{10}} \left[ \frac{1}{\sqrt{3}} \rel((H^L_+ + H^L_-)S^{L \ast}) - \frac{1}{\sqrt{15}} \rel((H^L_+ + H^L_-)D^{L \ast}_0  ) + \frac{1}{\sqrt{5}} \rel((D^L_+ + D^L_-)H^{L \ast}_0 ) \right] $  & "\\  \hline
 12  &   $P^1_2 \sqrt{2}\rel(Y^1_2)$ &  $-\frac{3}{5\sqrt{2}} \left[ \rel( (H^L_+ + H^L_-)H^{L \ast}_0)  + \sqrt{\frac{5}{3}} \rel ((D^L_+ + D^L_-) S^{L \ast})  + \frac{5}{7 \sqrt{3}} \rel((D^L_+ + D^L_-) D^{L\ast}_0) \right] $ & " \\  \hline
 13  &   $P^1_3 \sqrt{2}\rel(Y^1_2)$ &  $-\frac{3}{5 \sqrt{7}} \left[2 \rel((D^L_+ + D^L_-)H^{L\ast}_0)  + \sqrt{3} \rel((H^L_+ + H^L_-)D^{L\ast}_0 ) \right] $ & " \\  \hline
 14  &   $P^1_4 \sqrt{2}\rel(Y^1_2)$ &  $- \frac{3}{7} \rdpdmdz$  & " \\  \hline
 15  &   $P^1_1 \sqrt{2}\img(Y^1_2)$ &  $\frac{3}{\sqrt{2}}\left[ \frac{1}{\sqrt{15}} \img((H^L_+ - H^L_-)S^{L\ast}) + \frac{1}{5} \img((D^L_+ -D^L_-)H^{L\ast}_0) - \frac{1}{5 \sqrt{3}}  \img((H^L_+ -H^L_-)D^{L\ast}_0) \right]  $  & " \\  \hline
 16  &   $P^1_2 \sqrt{2}\img(Y^1_2)$ &  $ \frac{3}{\sqrt{2}} \left[ \frac{1}{7 \sqrt{3}} \img((D^L_+ -D^L_-)D^{L\ast}_0)  + \frac{1}{5} \img((H^L_+ -H^L_-)H^{L\ast}_0)  + \frac{1}{\sqrt{15}} \img((D^L_+ -D^L_-)S^{L\ast})   \right] $  & " \\  \hline
 17  &   $P^1_3 \sqrt{2}\img(Y^1_2)$ &  $\frac{3}{5 \sqrt{7}} \left[ 2 \img((D^L_+ - D^L_-)H^{L\ast}_0)  + \sqrt{3} \img((H^L_+ - H^L_-)D^{L\ast}_0) \right]  $   & " \\  \hline
 18  &   $P^1_4 \sqrt{2}\img(Y^1_2)$ &  $\frac{3}{7} \img((D^L_+ -D^L_-)D^{L\ast}_0)$  & " \\  \hline
 19  &   $P^0_0 \sqrt{2}\rel(Y^2_2)$ &  $-\sqrt{\frac{3}{5}} \left[ \rel(H^L_+ H^{L\ast}_-)   + \rel(D^L_+ D^{L\ast}_-) \right] $  & " \\  \hline
 20  &   $P^0_1 \sqrt{2}\rel(Y^2_2)$ &  $- \frac{3}{5} \left[ \rel(H^L_+ D^{L\ast}_-)   + \rel(D^L_+ H^{L\ast}_-) \right] $  & " \\  \hline
 21  &   $P^0_2 \sqrt{2}\rel(Y^2_2)$ &  $\sqrt{3} \left[ - \frac{1}{7} \rel(D^L_+ D^{L\ast}_-)   + \frac{1}{5} \rel(H^L_+ H^{L\ast}_-) \right] $  & " \\  \hline
 22  &   $P^0_3 \sqrt{2}\rel(Y^2_2)$ &  $\frac{3}{5} \sqrt{ \frac{3}{7}} \left[ \rel(H^L_+ D^{L\ast}_-)   + \rel(D^L_+ H^{L\ast}_-) \right] $  & " \\  \hline
 23  &   $P^0_4 \sqrt{2}\rel(Y^2_2)$ &  $\frac{4}{7} \sqrt{ \frac{3}{5}}  \rel(D^L_+ D^{L\ast}_-) $ & " \\  \hline
 24  &   $P^0_0 \sqrt{2}\img(Y^2_2)$ &  $\sqrt{\frac{3}{5}} \left[ \img(H^L_+ H^{L\ast}_-) + \img(D^L_+ D^{L\ast}_-) \right] $   & " \\  \hline
 25  &   $P^0_1 \sqrt{2}\img(Y^2_2)$ &  $\frac{3}{5} \img(H^L_+ D^{L\ast}_- + D^L_+ H^{L\ast}_-)  $  & " \\ \hline
 26  &   $P^0_2 \sqrt{2}\img(Y^2_2)$ &  $ \sqrt{3} \left[\frac{1}{7} \img(D^L_+ D^{L\ast}_-)   - \frac{1}{5} \img(H^L_+ H^{L\ast}_-)\right] $  & " \\ \hline
 27  &   $P^0_3 \sqrt{2}\img(Y^2_2)$ &  $-\frac{3}{5} \sqrt{ \frac{3}{7}}  \img(H^L_+ D^{L\ast}_- + D^L_+ H^{L\ast}_-)  $  & " \\ \hline
 28  &   $P^0_4 \sqrt{2}\img(Y^2_2)$ &  $-\frac{4}{7} \sqrt{\frac{3}{5}}  \img(D^L_+ D^{L\ast}_-) $   & " \\ \hline \hline
 29  &   $P^0_0 Y^0_1$     &  $\frac{\sqrt{3}}{2} \left[ (\hmsq - \hpsq) + (\dmsq - \dpsq) \right]$  & - ($L \to R$) \\ \hline
 30  &   $P^0_1 Y^0_1$     &  $\frac{3}{\sqrt{5}} \rel( H^L_- D^{L\ast}_- - H^L_+ D^{L\ast}_+ ) $ & " \\ \hline
 31  &   $P^0_2 Y^0_1$     &  $\frac{3}{2\sqrt{15}} \left[ \frac{5}{7}(\dmsq - \dpsq) - (\hmsq - \hpsq) \right]$ & " \\ \hline
 32  &   $P^0_3 Y^0_1$     &  $-\frac{9}{\sqrt{105}}  \rel(H^L_- D^{L\ast}_-   -H^L_+ D^{L\ast}_+ ) $ & " \\ \hline
 33  &   $P^0_4 Y^0_1$     &  $-\frac{2\sqrt{3}}{7} (\dmsq - \dpsq)$  & " \\ \hline
 34  &   $P^1_1 \sqrt{2}\rel(Y^1_1)$   & $\sqrt{\frac{3}{10}} \left[ \sqrt{5} \rel((H^L_+ -H^L_-)S^{L \ast})  + \sqrt{3} \rel((D^L_+ -D^L_-)H^{L\ast}_0)  - \rel((H^L_+ -H^L_-)D^{L \ast}_0) \right]$  & " \\ \hline
 35  &   $P^1_2 \sqrt{2}\rel(Y^1_1)$   & $ \frac{3}{\sqrt{2}}\left[ \frac{1}{\sqrt{5}} \rel((H^L_+ -H^L_-)H^{L \ast}_0)  + \frac{1}{\sqrt{3}} \rel((D^L_+ -D^L_-)S^{L\ast})  + \frac{5}{21} \sqrt{\frac{3}{5}} \rel((D^L_+ -D^L_-)D^{L \ast}_0 ) \right] $  & " \\ \hline
 36  &   $P^1_3 \sqrt{2}\rel(Y^1_1)$   & $ \frac{3}{\sqrt{35}} \left[ 2 \rel((D^L_+ -D^L_-)H^{L \ast}_0)  + \sqrt{3} \rel((H^L_+ -H^L_-)D^{L\ast}_0) \right]$  & " \\ \hline
 37  &   $P^1_4 \sqrt{2}\rel(Y^1_1)$   & $\frac{3}{7} \sqrt{5} \rel((D^L_+ -D^L_-)D^{L \ast}_0 ) $  & " \\ \hline
 38  &   $P^1_1 \sqrt{2}\img(Y^1_1)$   & $-\sqrt{\frac{3}{10}} \left[ \sqrt{5} \img ( (H^L_+ +H^L_-)S^{L\ast}) + \sqrt{3} \img((D^L_+ +D^L_-)H^{L \ast}_0) - \img((H^L_+ +H^L_-)D^{L \ast}_0)  \right]  $  & " \\ \hline
 39  &   $P^1_2 \sqrt{2}\img(Y^1_1)$   & $ -\sqrt{\frac{3}{10}} \left[ \sqrt{3} \img((H^L_+ +H^L_-)H^{L \ast}_0)  + \sqrt{5} \img((D^L_+ +D^L_-)S^{L\ast})  + \frac{5}{7} \img((D^L_+ +D^L_-)D^{L \ast}_0 )\right] $  & " \\ \hline
 40  &   $P^1_3 \sqrt{2}\img(Y^1_1)$   & $ -\frac{3}{\sqrt{35}}\left[ 2\img((D^L_+ +D^L_-)H^{L \ast}_0)  + \sqrt{3} \img((H^L_+ +H^L_-)D^{L\ast}_0)\right]$   & " \\ \hline
 41  &   $P^1_4 \sqrt{2}\img(Y^1_1)$   & $-\frac{3}{7} \sqrt{5 } \img((D^L_+ +D^L_-)D^{L\ast}_0) $  & " \\ \hline
\end{tabular}
\caption{The helicity-basis moments of the 41 orthonormal angular functions $f_i(\Omega)$ in Eq.~\ref{eqn:vector_moments}.}
\label{table:spd_mom_hel}
\end{center}
\end{table*}

\begin{table*}
\begin{center}
\renewcommand{\arraystretch}{1.31}
\begin{tabular}{c|c|c|c} 
 $i$    &   $f_i(\Omega)$             & $\Gamma^{L, {\rm tr}}_i(\qsq)$ & $\eta^{L\to R}_i$  \\ \hline \hline
 1   &   $P^0_0 Y^0_0$     &  $\left[ \hzsq + \hpasq + \hpesq + \ssq + \dzsq + \dpasq + \dpesq\right]$ & + ($L \to R$)\\ \hline 
 2   &   $P^0_1 Y^0_0$     &  $2\left[\frac{2}{\sqrt{5}} \rhzdz + \rshz + \sqrt{\frac{3}{5}}  \rel( H^L_\parallel D^{L\ast}_\parallel + H^L_\perp D^{L\ast}_\perp  )\right]$ & " \\ \hline 
 3   &   $P^0_2 Y^0_0$     &  $\frac{\sqrt{5}}{7}$ (\dpasq + \dpesq) - $\frac{1}{\sqrt{5}}$ (\hpasq + \hpesq) + $\frac{2}{\sqrt{5}}$ \hzsq  + $\frac{10}{7\sqrt{5}}$ \dzsq + $2$ \rsdz & " \\  \hline
 4   &   $P^0_3 Y^0_0$     &  $\frac{6}{\sqrt{35}} \left[ - \rel(H^L_\parallel D^{L\ast}_\parallel +  H^L_\perp D^{L\ast}_\perp)  + \sqrt{3} \rhzdz  \right]$ & "\\  \hline
 5   &   $P^0_4 Y^0_0$     &  $\frac{2}{7} \left[ -2 (\dpasq + \dpesq) + 3 \dzsq \right] $ & "\\  \hline
 6   &   $P^0_0 Y^0_2$     &  $\frac{1}{2 \sqrt{5}} \left[ (\dpasq + \dpesq) + (\hpasq + \hpesq) - 2 \ssq - 2 \dzsq - 2 \hzsq \right]$ & " \\  \hline
 7   &   $P^0_1 Y^0_2$     &  $\left[ \frac{\sqrt{3}}{5} \rel(H^L_\parallel D^{L\ast}_\parallel  + H^L_\perp D^{L\ast}_\perp) - \frac{2}{\sqrt{5}} \rel(S^L H^{L\ast}_0)  - \frac{4}{5} \rel(H^L_0 D^{L\ast}_0)\right] $ & "  \\ \hline
 8   &   $P^0_2 Y^0_2$     &  $ \left[ \frac{1}{14} (\dpasq + \dpesq) - \frac{2}{7} \dzsq - \frac{1}{10} (\hpasq + \hpesq) - \frac{2}{5} \hzsq - \frac{2}{\sqrt{5}} \rsdz \right]$ & "  \\  \hline
 9   &   $P^0_3 Y^0_2$     &  $ - \frac{3}{5 \sqrt{7}} \left[ \rel( H^L_\parallel D^{L \ast}_\parallel + H^L_\perp D^{L \ast}_\perp) + 2 \sqrt{3} \rel(H^L_0 D^{L \ast}_0 ) \right] $ & "\\  \hline
 10  &   $P^0_4 Y^0_2$     &  $ -\frac{2}{7 \sqrt{5}}  \left[ \dpasq + \dpesq + 3 \dzsq \right] $ & "  \\  \hline
 11  &   $P^1_1 \sqrt{2}\rel(Y^1_2)$ &  $-\frac{3}{\sqrt{10}} \left[ \sqrt{\frac{2}{3}} \rel(H^L_\parallel S^{L \ast}) - \sqrt{\frac{2}{15}} \rel(H^L_\parallel D^{L \ast}_0  ) + \sqrt{\frac{2}{5}} \rel(D^L_\parallel H^{L \ast}_0 ) \right] $  & "\\  \hline
 12  &   $P^1_2 \sqrt{2}\rel(Y^1_2)$ &  $-\frac{3}{5} \left[ \rel( H^L_\parallel H^{L \ast}_0)  + \sqrt{\frac{5}{3}} \rel (D^L_\parallel S^{L \ast})  + \frac{5}{7 \sqrt{3}} \rel(D^L_\parallel D^{L\ast}_0) \right] $ & " \\  \hline
 13  &   $P^1_3 \sqrt{2}\rel(Y^1_2)$ &  $-\frac{6}{5 \sqrt{14}} \left[2 \rel(D^L_\parallel H^{L\ast}_0)  + \sqrt{3} \rel(H^L_\parallel D^{L\ast}_0 ) \right] $ & " \\  \hline
 14  &   $P^1_4 \sqrt{2}\rel(Y^1_2)$ &  $- \frac{6}{7\sqrt{2}} \rel(D^L_\parallel D^{L\ast}_0)$  & " \\  \hline
 15  &   $P^1_1 \sqrt{2}\img(Y^1_2)$ &  $3 \left[ \frac{1}{\sqrt{15}} \img(H^L_\perp S^{L\ast}) + \frac{1}{5} \img(D^L_\perp H^{L\ast}_0) - \frac{1}{5 \sqrt{3}}  \img(H^L_\perp D^{L\ast}_0) \right]  $  & " \\  \hline
 16  &   $P^1_2 \sqrt{2}\img(Y^1_2)$ &  $ 3\left[ \frac{1}{7 \sqrt{3}} \img(D^L_\perp D^{L\ast}_0)  + \frac{1}{5} \img(H^L_\perp H^{L\ast}_0)  + \frac{1}{\sqrt{15}} \img(D^L_\perp S^{L\ast})   \right] $  & " \\  \hline
 17  &   $P^1_3 \sqrt{2}\img(Y^1_2)$ &  $\frac{6}{5 \sqrt{14}} \left[ 2 \img(D^L_\perp H^{L\ast}_0)  + \sqrt{3} \img(H^L_\perp D^{L\ast}_0) \right]  $   & " \\  \hline
 18  &   $P^1_4 \sqrt{2}\img(Y^1_2)$ &  $\frac{6}{7\sqrt{2}} \img(D^L_\perp D^{L\ast}_0)$  & " \\  \hline
 19  &   $P^0_0 \sqrt{2}\rel(Y^2_2)$ &  $-\frac{3}{2\sqrt{15}} \left[ (\hpasq - \hpesq) + (\dpasq - \dpesq) \right] $  & " \\  \hline
 20  &   $P^0_1 \sqrt{2}\rel(Y^2_2)$ &  $-\frac{3}{5} \left[ \rel(H^L_\parallel D^{L\ast}_\parallel)   - \rel(D^L_\perp H^{L\ast}_\perp) \right] $  & " \\  \hline
 21  &   $P^0_2 \sqrt{2}\rel(Y^2_2)$ &  $\frac{\sqrt{3}}{2} \left[ - \frac{1}{7} (\dpasq - \dpesq)   + \frac{1}{5} ( \hpasq - \hpesq ) \right] $  & " \\  \hline
 22  &   $P^0_3 \sqrt{2}\rel(Y^2_2)$ &  $\frac{3}{5} \sqrt{ \frac{3}{7}} \left[ \rel(H^L_\parallel D^{L\ast}_\parallel)   - \rel(D^L_\perp H^{L\ast}_\perp) \right] $  & " \\  \hline
 23  &   $P^0_4 \sqrt{2}\rel(Y^2_2)$ &  $\frac{2}{7} \sqrt{ \frac{3}{5}}  (\dpasq - \dpesq) $ & " \\  \hline
 24  &   $P^0_0 \sqrt{2}\img(Y^2_2)$ &  $\sqrt{\frac{3}{5}} \left[ \img(H^L_\perp H^{L\ast}_\parallel) + \img(D^L_\perp D^{L\ast}_\parallel) \right] $   & " \\  \hline
 25  &   $P^0_1 \sqrt{2}\img(Y^2_2)$ &  $\frac{3}{5} \img(  H^L_\perp D^{L\ast}_\parallel +  D^L_\perp H^{L\ast}_\parallel )  $  & " \\ \hline
 26  &   $P^0_2 \sqrt{2}\img(Y^2_2)$ &  $ \sqrt{3} \left[\frac{1}{7} \img(D^L_\perp D^{L\ast}_\parallel)   - \frac{1}{5} \img(H^L_\perp H^{L\ast}_\parallel)\right] $  & " \\ \hline
 27  &   $P^0_3 \sqrt{2}\img(Y^2_2)$ &  $-\frac{3}{5} \sqrt{ \frac{3}{7}}  \img(D^L_\perp H^{L\ast}_\parallel + H^L_\perp D^{L\ast}_\parallel)  $  & " \\ \hline
 28  &   $P^0_4 \sqrt{2}\img(Y^2_2)$ &  $-\frac{4}{7} \sqrt{\frac{3}{5}}  \img(D^L_\perp D^{L\ast}_\parallel) $   & " \\ \hline \hline
 29  &   $P^0_0 Y^0_1$     &  $-\sqrt{3}\left[ \rel(H^L_\perp H^{L\ast}_\parallel) + \rel(D^L_\perp D^{L\ast}_\parallel) \right]$  & - ($L \to R$) \\ \hline
 30  &   $P^0_1 Y^0_1$     &  $-\frac{3}{\sqrt{5}} \rel( H^L_\perp D^{L\ast}_\parallel + H^L_\parallel D^{L\ast}_\perp ) $ & " \\ \hline
 31  &   $P^0_2 Y^0_1$     &  $-\frac{3}{\sqrt{15}} \left[ \frac{5}{7} \rel(D^L_\perp D^{L\ast}_\parallel) - \rel(H^L_\perp H^{L\ast}_\parallel)  \right]$ & " \\ \hline
 32  &   $P^0_3 Y^0_1$     &  $\frac{9}{\sqrt{105}}  \rel(H^L_\perp D^{L\ast}_\parallel  + H^L_\parallel D^{L\ast}_\perp ) $ & " \\ \hline
 33  &   $P^0_4 Y^0_1$     &  $\frac{4\sqrt{3}}{7} \rel(D^L_\perp D^{L\ast}_\parallel)$  & " \\ \hline
 34  &   $P^1_1 \sqrt{2}\rel(Y^1_1)$   & $\sqrt{\frac{3}{5}} \left[ \sqrt{5} \rel(H^L_\perp S^{L \ast})  + \sqrt{3} \rel(D^L_\perp H^{L\ast}_0)  - \rel(H^L_\perp D^{L \ast}_0) \right]$  & " \\ \hline
 35  &   $P^1_2 \sqrt{2}\rel(Y^1_1)$   & $ 3 \left[ \frac{1}{\sqrt{5}} \rel(H^L_\perp H^{L \ast}_0)  + \frac{1}{\sqrt{3}} \rel(D^L_\perp S^{L\ast})  + \frac{5}{21} \sqrt{\frac{3}{5}} \rel(D^L_\perp D^{L \ast}_0 ) \right] $  & " \\ \hline
 36  &   $P^1_3 \sqrt{2}\rel(Y^1_1)$   & $ \frac{6}{\sqrt{70}} \left[ 2 \rel(D^L_\perp H^{L \ast}_0)  + \sqrt{3} \rel(H^L_\perp D^{L\ast}_0) \right]$  & " \\ \hline
 37  &   $P^1_4 \sqrt{2}\rel(Y^1_1)$   & $\frac{3 \sqrt{10}}{7} \rel(D^L_\perp D^{L \ast}_0 ) $  & " \\ \hline
 38  &   $P^1_1 \sqrt{2}\img(Y^1_1)$   & $-\sqrt{\frac{3}{5}} \left[ \sqrt{5} \img ( H^L_\parallel S^{L\ast}) + \sqrt{3} \img(D^L_\parallel H^{L \ast}_0) - \img(H^L_\parallel D^{L \ast}_0)  \right]  $  & " \\ \hline
 39  &   $P^1_2 \sqrt{2}\img(Y^1_1)$   & $ -\sqrt{\frac{3}{5}} \left[ \sqrt{3} \img(H^L_\parallel H^{L \ast}_0)  + \sqrt{5} \img(D^L_\parallel S^{L\ast})  + \frac{5}{7} \img(D^L_\parallel D^{L \ast}_0 )\right] $  & " \\ \hline
 40  &   $P^1_3 \sqrt{2}\img(Y^1_1)$   & $ -6\sqrt{\frac{1}{70}}\left[ 2\img(D^L_\parallel H^{L \ast}_0)  + \sqrt{3} \img(H^L_\parallel D^{L\ast}_0)\right]$   & " \\ \hline
 41  &   $P^1_4 \sqrt{2}\img(Y^1_1)$   & $-\frac{3}{7} \sqrt{10} \img(D^L_\parallel D^{L\ast}_0) $  & " \\ \hline
\end{tabular}
\caption{The transversity-basis moments of the 41 orthonormal angular functions $f_i(\Omega)$ in Eq.~\ref{eqn:vector_moments}.}
\label{table:spd_mom_trans}
\end{center}
\end{table*}

\section{The two-fold ambiguity}
\label{sec:ambiguities}

As mentioned in the introduction, the full differential rate does not uniquely determine the helicity amplitudes. The ambiguities in the solutions arise from the informtion loss in summing over the final lepton spins. A detailed study of these ambiguities is beyond the scope of this work. However, we point out one particular case.

Using the identities $d^J_{\lambda,0} \equiv (-1)^\lambda d^J_{-\lambda,0}$ and $d^1_{\lambda,\eta} \equiv -(-1)^\lambda d^1_{-\lambda,-\eta}$ for $\eta = \pm 1$ and $\lambda \in \{0,\pm 1\}$, the expression in Eq.~\ref{eqn:rate_M2_XP1P2} is seen to be invariant under the following global transformation:
\begin{align}
\label{eqn:twofoldamb}
\mathcal{H}^{\eta,J}_\lambda \to \left(\mathcal{H}^{-\eta,J}_{-\lambda}\right)^\ast.
\end{align}
We note here again that $\eta = +1 (-1)$ denotes the RH(LH) component on the leptonic side. For the electromagnetic $c \bar{c} \to \ell^+ \ell^-$ decays, the LH and RH amplitudes are equal and Eq.~\ref{eqn:twofoldamb} represents the two-fold ambiguity~\cite{babar_verderi2005} in the determination of $\beta$ and $\beta_s$ from $B \to J/\psi K^\ast$ and  $B_s \to J/\psi \phi$, respectively.

\section{Analysis formalism}

\subsection{No background case}

\subsubsection{Method of Moments (MOM)}

Assume a generic rate function constructed out of a set of orthonormal basis functions $f_i(\Omega)$:
\begin{align}
\label{eqn:pdf_defn}
\displaystyle \frac{dN}{d\Omega} \equiv g(\Omega) = \displaystyle \sum_i b_i f_i(\Omega),
\end{align}
where the aim is to determine the moments $b_i$. We define a detector efficiency function $\epsilon(\Omega)$, and the normalization integrals
\begin{align}     
 E_{( i,j,\cdots,n)} &= \int \epsilon (\Omega) \left[ f_i(\Omega) f_j(\Omega) \cdots f_n(\Omega)\right] d \Omega \nonumber \\
                    &= \displaystyle \frac{\Phi}{N^{\rm MC}_{\rm gen}} \left[\sum_{k=1}^{N^{\rm MC}_{\rm acc} } f_i(\Omega_k) f_j(\Omega_k) \cdots f_n(\Omega_k)\right],   
\label{eqn:norm_ints}
\end{align}      
that are calculated numerically with $N^{\rm MC}_{\rm gen}$ Monte Carlo (MC) events generated flat in $d\Omega$, and $N^{\rm MC}_{\rm acc}$ accepted events that survive after the detector efficiency is taken into account. Also, $\Phi = \int d \Omega$ is the total phase-space element.

The measured moments from the data are
\begin{align}
\tilde{b}_i &\equiv \displaystyle \sum_{k=1}^{ N^{\rm data}} f_i(\Omega_k) = \int f_i(\Omega) \epsilon(\Omega) \frac{dN}{d\Omega} d\Omega = E_{ij} b_j 
\end{align}
from which, the efficiency-corrected true moments can be calculated as
\begin{align}
b_i &= \displaystyle (E^{-1})_{ij} \tilde{b}_j.
\end{align}
Likewise, the measured covariance matrix of the moments is estimated as
\begin{align}
\label{eqn:C_tilde}
\tilde{C}_{ij} = 
\displaystyle \sum_{k=1}^{ N^{\rm data}} f_i(\Omega_k) f_j(\Omega_k) &= \int f_i(\omega) f_j(\omega)\epsilon(\Omega) \frac{dN}{d\Omega} d\Omega \nonumber \\ &= E_{ijk} b_k,
\end{align}
and the covariance matrix of the acceptance corrected moments are
\begin{align}
C_{ij} = (E^{-1})_{ik} \tilde{C}_{kl} (E^{-1})_{jl}.
\end{align}

In the next step of the method of moments (MOM), if the moments functions are parameterized by a set of parameters $\vec{\alpha}$ in some physics-motivated model as $b_i(\vec{\alpha})$, the values of the $\vec{\alpha}$ can be obtained by minimizing the $\chi^2$ function
\begin{align}
\chi^2 = \sum _{ij} [b_i - b_i(\vec{\alpha})]\; [C^{-1}]_{ij} \; [b_j - b_j(\vec{\alpha})]
\label{eqn:chi2_minimization}
\end{align}

\subsubsection{Unbinned maximum-likelihood fits (UML)}

In the equivalent unbinned maximum-likelihood (UML) method, the efficiency incorporated pdf is
\begin{align}
g'(\Omega,\vec{\alpha}) = \frac{\displaystyle \sum_i b_i(\vec{\alpha}) f_i(\Omega) \epsilon(\Omega)}{\displaystyle \sum_i b_i(\vec{\alpha}) E_i}.
\end{align}
The likelihood function to maximize is
\begin{align}
\mathcal{L}(\vec{\alpha}) = \prod_{k=1}^{N^{\rm data}} g'(\Omega_k,\vec{\alpha}),
\end{align}
which leads to the negative log-likelihood (NLL) to minimize as
\begin{align}
-2 \ln(\mathcal{L}(\vec{\alpha})) &= 2N^{\rm data} \ln \left[\displaystyle \sum_i b_i(\vec{\alpha}) E_i \right] \nonumber \\
   & \hspace{0.7cm}- 2 \displaystyle \sum_{k=1}^{N^{\rm data}} \ln \left[\displaystyle \sum_i b_i(\vec{\alpha}) f_i(\Omega_k)  \right].
\label{eqn:nll_nobkg}
\end{align}

\subsection{\label{sec:toy_studies}Studies with toy Monte Carlo}

To validate the above expressions, we consider a simple rate expression for toy studies:
\begin{align}
\frac{dN}{d\theta} \equiv g(\theta,\alpha,\beta) = \frac{N}{\pi + 2 \beta} ( 1 + \alpha \cos \theta + \beta \sin \theta)
\label{eqn:g_toy}
\end{align}
with $\theta \in [0,\pi]$ and $\{\alpha,\beta\}$ being the target parameters to be determined. The total number of events, $N$, is a nuisance parameter for the moment. The orthonormal basis functions are  
\begin{subequations}
\label{eqn:f_toy}
\begin{align}
f_1 &= \frac{1}{\sqrt{\pi}} \\
f_2 &= \frac{\cos \theta}{\sqrt{\pi/2}} \\
f_3 &= \frac{\sin \theta - 2/ \pi}{\sqrt{\pi /2 - 4/\pi}},
\end{align}
\end{subequations}
and the corresponding moments
\begin{subequations}
\label{eqn:b_toy}
\begin{align}
b_1 &= \frac{N}{\sqrt{\pi}} \\
b_2 &= \frac{N\alpha\sqrt{\pi / 2}}{\pi + 2 \beta} \\
b_3 &= \frac{N\beta}{\pi + 2 \beta} \sqrt{\pi /2 - 4/\pi}.
\end{align}
\end{subequations}

Without any loss of generality, we model the detector efficiency as the three sets of functions given in Table~\ref{table:eff_fcns}. Specifically, we note that Set~III incorporates a ``hole'' in the detector around $\theta = \pi /2$, where the efficiency drops to zero.  

\begin{table}
\begin{center}
\begin{tabular}{ c|c  } 
Set\;& efficiency $\epsilon(\theta)$ \\ \hline \hline 
I & $(1 + \sin 2 \theta)/2$ \\ \hline
II & $(1+ \cos^3 \theta)/2$ \\ \hline
III & $(1+ \cos^3 \theta)/2$; 0 for $|\theta - \frac{\pi}{2}| < 0.1$
\end{tabular}
\caption{The different efficiency functions used in the toy studies of Sec.~\ref{sec:toy_studies}.}
\label{table:eff_fcns}
\end{center}
\end{table}

\begin{figure}
\begin{center}
\includegraphics[width=3in]{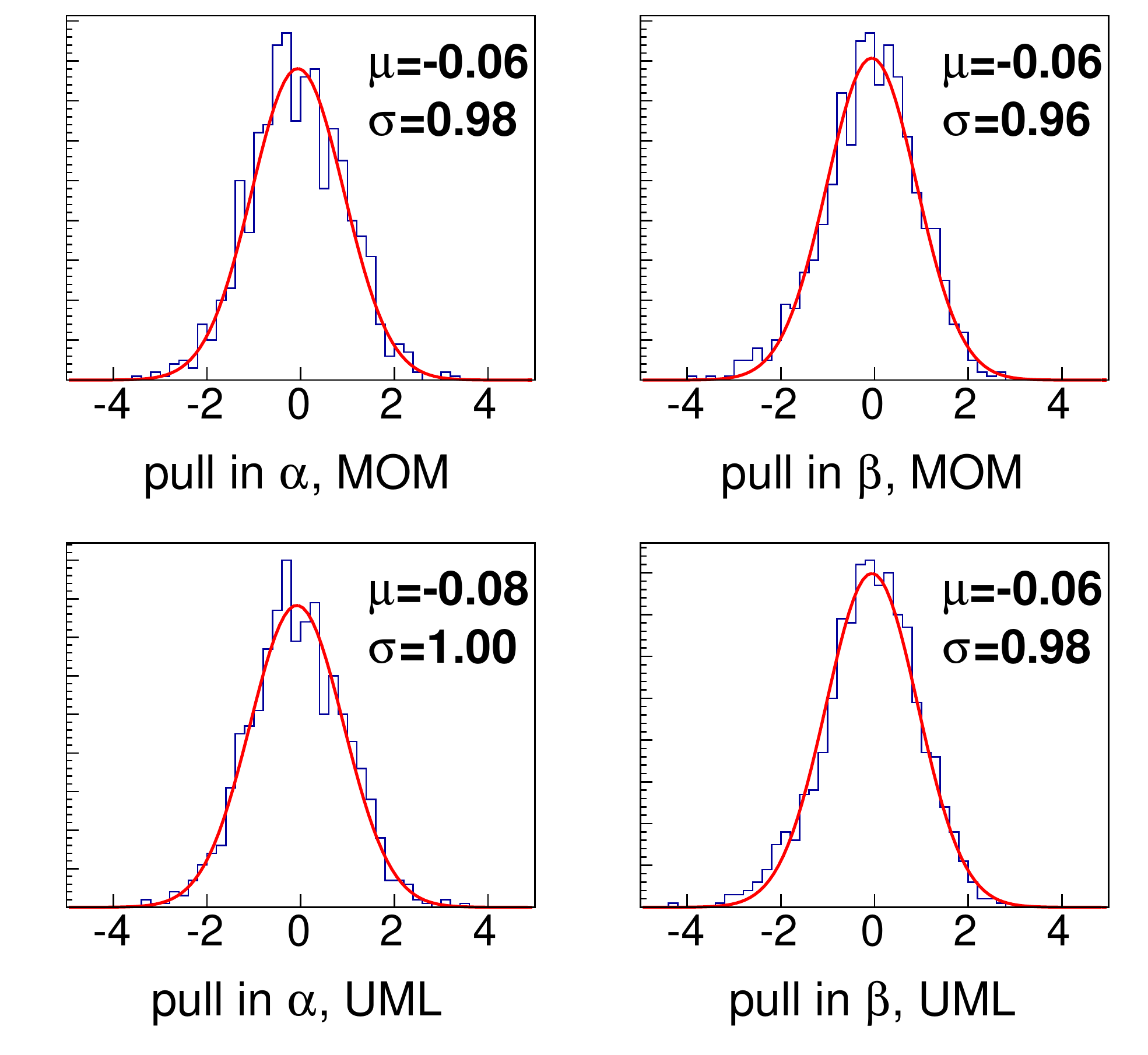}
\caption[]{\label{fig:toy_eff_nobkg_set3} (Color online) Pull distributions from a thousand toy samples of fits to Eq.~\ref{eqn:g_toy} with the efficiency function as Set~III in Table~\ref{table:eff_fcns}. The upper and lower plots use the MOM and UML techniques, respectively. No background is included.}
\end{center}
\end{figure}

Figure~\ref{fig:toy_eff_nobkg_set3} shows the pull distributions from fits to a thousand toy samples generated according to Eq.~\ref{eqn:g_toy} and the efficiency function as Set~III in Table~\ref{table:eff_fcns}.

\subsection{Incorporating background}

\begin{figure}
\begin{center}
\includegraphics[width=3in]{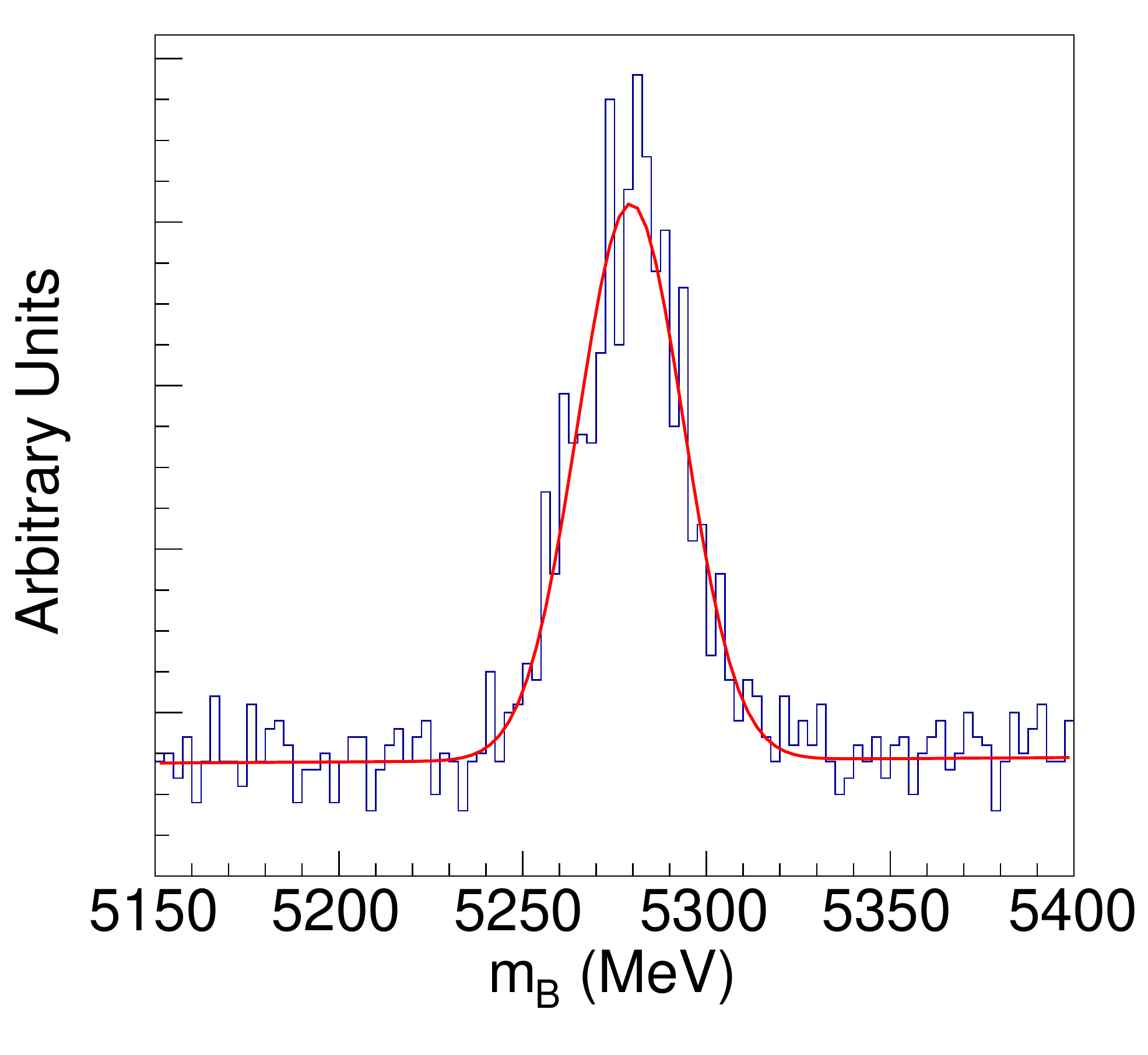}
\caption[]{\label{fig:sig_bkgd_fit_pur4} (Color online) A toy sample incorporating both efficiency and background effects used in the validation study. A fit to the profile is shown as well.}
\end{center}
\end{figure}

Next, to incorporate background, we assume that there is a discriminating variable $m$, un-correlated with the angular variables $\Omega$. Let there be $N^{\rm data}$ events (signal and background combined) in the ``signal region'' in the variable $m$, and $N^{\rm b}$ events in a suitably defined ``sideband region'', containing pure background events. Also, let $\tilde{n}_{\rm b}$ be the estimated background under the signal peak in the ``signal region'', obtained from a signal-background separation fit in the variable $m$. 

Independent toy sample sets with different purity levels were generated. Figure~\ref{fig:sig_bkgd_fit_pur4} shows the case for a toy sample with the discriminating variable $m \equiv m_b$ representative of the $B$ mass. The signal lineshape is a Gaussian while the background is constant. The ``signal region'' is chosen as $\pm 2\sigma$ around the mean, as obtained from the signal-background fit. The low and high sideband regions are taken as $m_B < 5200$~MeV and $m_B> 5360$~MeV, respectively. The background is generated flat in $m_B$ and $\theta$, but folded with the relevant efficiency functions in Table~\ref{table:eff_fcns}.

The ``pseudo-likelihood'' $\mathcal{L}'$ is then defined by assigning negative weights to the events in the sideband region:
\begin{align}
-2 \ln(\mathcal{L}'(\vec{\alpha})) &= 2 (N^{\rm data}-\tilde{n}_b) \ln \left[\displaystyle \sum_i b_i(\vec{\alpha}) E_i \right] \nonumber \\
   & \hspace{0.7cm}- 2 \displaystyle \sum_{k=1}^{N^{\rm data}} \ln \left[\displaystyle \sum_i b_i(\vec{\alpha}) f_i(\Omega_k)  \right] \nonumber \\
   & \hspace{0.7cm}+ 2 x \displaystyle \sum_{k=1}^{N^{\rm b}} \ln \left[\displaystyle \sum_i b_i(\vec{\alpha}) f_i(\Omega_k)  \right],
\label{eqn:nll_bkg}
\end{align}
where $x = \tilde{n}_{\rm b}/N_{\rm b}$ is a scale factor relating the background level under the signal, to that in the side-band.

Following the derivation in Refs.~\cite{babar_verderi2005,babar_verderi2007}, the covariance matrix from minimizing the pseudo-likelihood function in Eq.~\ref{eqn:nll_bkg} has to be modified to yield the true covariance matrix, $C^{\rm b}$, incorporating the additional uncertainties due to the background subtraction part as:
\begin{align}
C^{\rm b} &= C\left[ \mathbb{1} + \left\{ \tilde{n}_{\rm b} (1 + x) \mathcal{G} + N^2_{\rm b} \sigma^2_x \mathcal{H}\right\}C\right],
\label{eqn:corr_cov_mat}
\end{align}
where
\begin{align}
\mathcal{G}_{\lambda \mu} &= \frac{1}{N_{\rm b}} \sum_{k=1}^{N_{\rm b}} \left[ \frac{\partial \ln g'(\Omega_k,\vec{\alpha})}{\partial \alpha_\lambda}  \frac{\partial \ln g'(\Omega_k,\vec{\alpha})}{\partial \alpha_\mu} \right] \\
\mathcal{H}_{\lambda \mu} &= \left[  \frac{1}{N_{\rm b}} \sum_{k=1}^{N_{\rm b}} \frac{\partial \ln g'(\Omega_k,\vec{\alpha})}{\partial \alpha_\lambda} \right] \left[  \frac{1}{N_{\rm b}} \sum_{l=1}^{N_{\rm b}} \frac{\partial \ln g'(\Omega_l,\vec{\alpha})}{\partial \alpha_\mu} \right],
\end{align}
and $C$ is the covariance matrix returned by the {\tt HESSE} routine. Summing over repeated indices, the partial derivatives are explicitly
\begin{align}
\frac{\partial \ln g'(\Omega_k,\vec{\alpha})}{\partial \alpha_\lambda} &= \frac{\partial\; b_i(\vec{\alpha})}{\partial \alpha_\lambda} \left[ \frac{f_i(\Omega_k)}{b_j(\vec{\alpha})f_j(\Omega_k)} - \frac{E_i}{b_j(\vec{\alpha})E_j}\right]
\end{align}
and $\sigma_x$ is the uncertainty on the background scale factor $x$.

In the moments expansion method, the background-subtracted measured moments and the covariance matrix are estimated as
\begin{align}
\tilde{b}^{\rm b} &= \displaystyle \sum_{k=1}^{N^{\rm data}} f_i(\Omega_k) - x \displaystyle \sum_{k=1}^{N^{\rm b}} f_i(\Omega_k)\\
\tilde{C}^{\rm b}_{ij} &= \displaystyle \sum_{k=1}^{N^{\rm data}} f_i(\Omega_k) f_j(\Omega_k) + x^2 \displaystyle \sum_{k=1}^{N^{\rm b}} f_i(\Omega_k) f_j(\Omega_k)
\label{eqn:mom_bkg}
\end{align}

\begin{figure}
\begin{center}
\includegraphics[width=3in]{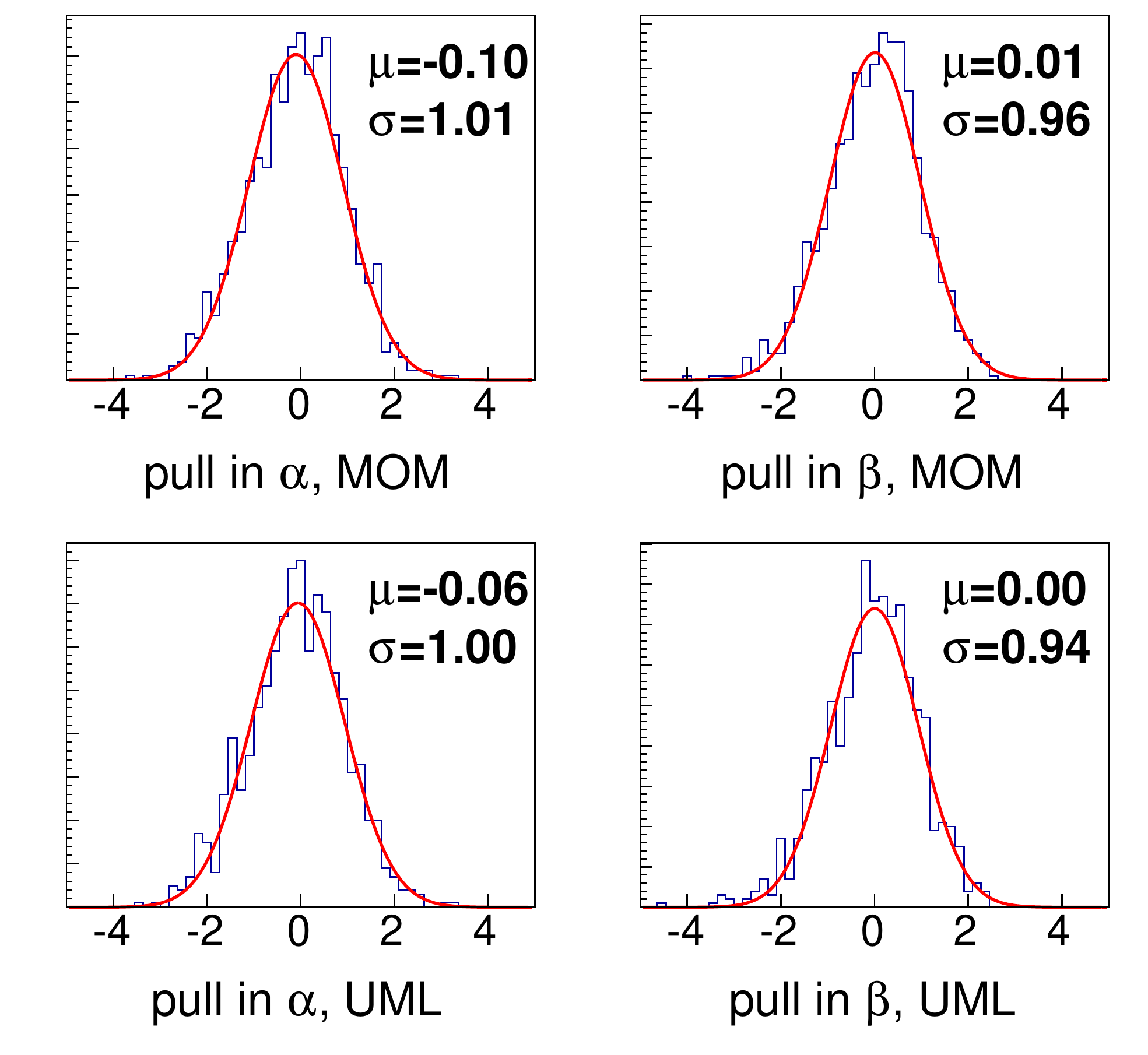}
\caption[]{\label{fig:toy_eff_set3_pur4} (Color online) Pull distributions from a thousand toy samples of fits to Eq.~\ref{eqn:g_toy} with the efficiency function as Set~III in Table~\ref{table:eff_fcns}. The upper and lower plots use the MOM and UML techniques, respectively. The samples incorporate a signal to background ratio corresponding to that in Fig.~\ref{fig:sig_bkgd_fit_pur4}.}
\end{center}
\end{figure}

The pull distributions from the MOM and ULM fits and the corresponding covariance matrices $\tilde{C}^{\rm b}$ and $C^{\rm b}$, respectively, are shown in Fig.~\ref{fig:toy_eff_set3_pur4}.

\subsection{Discussion}

We point here to some of the salient features of the MOM. The set of moments in Eq.~\ref{eqn:vector_moments} constitute a concise representation of all the angular information content in the entire dataset. The relations between the different moments and the amplitudes are ab initio not built in. These relations can be used as checks for understanding of the detector acceptance. They can also be incorporated during the model-dependent $\chi^2$ minimization fit as described by Eq.~\ref{eqn:chi2_minimization}. If the model-dependence is reliably known, the MOM and a direct UML fit give the same results, as we explicitly demostrated in Sec.~\ref{sec:toy_studies}. 

However, if the underlying physics model is unknown, the MOM can provide simple and model-independent confirmations of certain interesting physics features. For example, as pointed out in the introduction, a complex RH admixture $\epsR$ in the weak hadronic current leads to angular terms proportional to $\sin \chi$ in SL decays, that are absent in the SM. The presence of these terms in the data can be examined using any of the moments in Table~\ref{table:spd_mom_trans} corresponding to $Im(Y_l^m)$, where $m\neq0$. If the statistical significance of these moments are found to be high enough, this could constitute tension with the SM. 

Similarly, the observables $(|H_0|^2 + |S|^2)$, $|H_{\{\parallel,\perp\}}|^2$, $|D_{\{0.\parallel,\perp\}}|^2$ can be individually expressed in terms of the moments in Table~\ref{table:spd_mom_trans}. Therefore, if one is interested in the presence of a $D$-wave component under the $K^\ast(892)$ for $\Bbar \to \overline{K}^{\ast 0} \mu^- \mu^+$, this can be directly probed via the moments. In the absence of a $D$-wave component, the observables $|H_0|^2$ and $|S|^2$ can also be extracted directly from the moments, allowing an estimate of the $S$-wave fraction. For the observable $P'_5$~\cite{ffi_obs} that is predicted to be theoretically clean at low $\qsq$, the LHCb collaboration has recently observed~\cite{prl_lhcb_b2ksmumu} a $3.7 \sigma$ deviation from the SM. In the absence of non-$P$-wave components, this can be written in terms of the moments as:
\begin{align}
P'_5 = \displaystyle \sqrt{ \frac{5}{(\Gamma_1 + \sqrt{5} \Gamma_3)(\Gamma_1 - \sqrt{5} \Gamma_3/2)}}\; \Gamma_{35}.
\end{align}
The important point to note here is that no complicated multi-dimensional angular fit is required for any of these checks.

We would also like to comment on the use of the normalization integrals in Eq.~\ref{eqn:norm_ints} as opposed to analytic modeling of the efficiency function and reweighting of events by the inverse of the efficiency. The latter involves a complicated fit which can be unstable without due to local ``holes'' in the acceptance function. The normalization integrals, on the other hand, are found to be more robust under these situations.


\section{Summary}

In summary, we provide expressions for the full angular decay rate in $\overline{B}\to X\ell_1 \ell_2$ decays where the $\ell_2$ lepton can be either a charged $\{e,\mu \}$ or a neutrino. We considered the final state $X$ to include complex $S$-, $P$, and $D$-wave amplitudes. The rate expression is expanded in a basis of orthonormal moments functions and a procedure to extract the corresponding moments employing a counting measurement is desribed and validated. We expect the present work to be directly applicable to ongoing analyses at $\babar$ and LHCb.  

\appendix*

\section{Angle definitions}
\label{app:angles}

In this appendix we provide the explicit definition of the angles in terms of the 3-vectors. The definitions are equivalent to the GS definitions as explained in Sec.~\ref{sec:sign_conventions}.

We follow the convention adopted in App.~\ Ref.~\cite{ulrik_ambiguities} that the superscript on any 3-vector denotes the reference frame. For any ordered four-body final state $\B\to \{P_1,P_2,\ell_1,\ell_2\}$ where $P_{\{1,2\}}$ are pseudoscalars and $\ell_{\{1,2\}}$ are leptons, we define
\begin{subequations} 
\begin{align}
\vec{P}_{\ell_1 \ell_2} &= \vec{p}_{\ell_1} + \vec{p}_{\ell_2}\\
\vec{Q}_{\ell_1 \ell_2} &= \vec{p}_{\ell_1} - \vec{p}_{\ell_2}\\
\vec{P}_{P_1 P_2} &= \vec{p}_{\ell_1} + \vec{p}_{P_2}\\
\vec{Q}_{P_1 P_2} &= \vec{p}_{\ell_1} - \vec{p}_{P_2}.
\end{align}
\label{app:vecs1}
\end{subequations} 
The helicity angles are defined as
\begin{subequations} 
\begin{align}
\ctl &= - \displaystyle\frac{\vec{Q}^{\ell \ell}_{\ell_1 \ell_2} \cdot \vec{P}^{\ell \ell}_{P_1 P_2}}{ |\vec{Q}^{\ell \ell}_{\ell_1 \ell_2}| |\vec{P}^{\ell \ell}_{P_1 P_2}|} \\
\ctv &= -\displaystyle\frac{\vec{Q}^{PP}_{P_1 P_2} \cdot \vec{P}^{PP}_{\ell_1 \ell_2}}{ | \vec{Q}^{PP}_{P_1 P_2}   | |\vec{P}^{PP}_{\ell_1 \ell_2}|},
\end{align}
\label{app:vecs2}
\end{subequations} 
where $\ell \ell$ and $PP$ in the superscripts refer to the leptonic and hadronic rest frames. 

The normals to the two planes are defined as 
\begin{subequations} 
\begin{align}
\vec{N}_{\ell_1 \ell_2} &= -\vec{P}^B_{\ell_1 \ell_2}\times \vec{Q}^\B_{\ell_1 \ell_2}\\
\vec{N}_{P_1 P_2} &= \vec{P}^B_{P_1 P_2}\times \vec{Q}^B_{P_1 P_2},
\end{align}
\label{app:vecs3}
\end{subequations} 
and the dihedral angle between the planes is defined by
\begin{subequations} 
\begin{align}
\cos \chi &= - \displaystyle \frac{\vec{N}_{\ell_1 \ell_2} \cdot \vec{N}_{P_1 P_2}}{|\vec{N}_{\ell_1 \ell_2}| |\vec{N}_{P_1 P_2}|} \\
\sin \chi &= \displaystyle \left(  \frac{\vec{N}_{\ell_1 \ell_2} \times \vec{N}_{P_1 P_2}}{|\vec{N}_{\ell_1 \ell_2}| |\vec{N}_{P_1 P_2}|}  \right) \cdot \frac{ \vec{P}^B_{\ell_1 \ell_2}  }{ |\vec{P}^B_{\ell_1 \ell_2}|  }.
\end{align}
\label{app:vecs4}
\end{subequations}
For the $\overline{B}$ decay, our ordering is $\overline{B}^0 \to \{K^-\pi^+ \ell^- \ell^+\}$, leading to a single sign flip in $\ctl$ compared to the EWP theory convention, as was explained in Eq.~\ref{eqn:thetal_flip}.

For the CP conjugate decay $B \to \{\bar{P}_1 \bar{P}_2 \bar{\ell}_1 \bar{\ell}_2\}$, we perform the CP conjugation explicitly while maintaining the order. The same procedure using Eqs.~\ref{app:vecs1}-~\ref{app:vecs4} is applied to the CP conjugated system to yield the angles. This leads to a single sign flip in the angle $\chi$, as mentioned earlier in Sec.~\ref{sec:cp_conj}.

\begin{acknowledgments}
We thank Bill Dunwoodie for instigating interest in the utility of the moments technique and many helpful suggestions on the angular analysis formalism.
\end{acknowledgments}

\end{document}